\documentclass[journal]{IEEEtran}

\usepackage{packages/jabbrv} 
\usepackage[numbers,sort&compress,square]{natbib}

\usepackage{amsmath}
\usepackage{amsthm}
\usepackage{amsfonts}
\usepackage{amssymb}
\usepackage{bm}
\usepackage{bbm}

\usepackage{graphicx}

\usepackage{tabularx}
\usepackage{booktabs}
\usepackage{xcolor,colortbl} 
\usepackage{multirow}

\usepackage{ragged2e}

\usepackage{times}
\usepackage{enumitem}

\usepackage[hidelinks]{hyperref}
\usepackage{cleveref}

\usepackage{siunitx}
\usepackage{xspace}
\usepackage{array}
\usepackage{comment}
\usepackage{float}

\newcommand*{\etal}{ et al.\ \xspace}

\newcommand*{\ie}{i.e.,\xspace}

\newcommand{\etalcite}[1]{ et al.~\cite{#1}}
\newcommand*{\gccphat}{\mbox{GCC-PHAT}\xspace}
\newcommand{\aname}[1]{\emph{#1}}

\newcommand{\mbf}{\mathbf}

\newcommand{\T}{\mathsf{T}}
\renewcommand{\H}{\mathsf{H}}
\newcommand{\E}{\mathbb{E}}
\newcommand{\ER}{$\text{ER}_{\le \SI{20}{\degree}}$\xspace}
\newcommand{\Fone}{$\text{F}_{\le \SI{20}{\degree}}$\xspace}
\newcommand{\LE}{$\text{LE}_\text{CD}$\xspace}
\newcommand{\LR}{$\text{LR}_\text{CD}$\xspace}
\newcommand{\Eseld}{$\mathcal{E}_\text{SELD}$\xspace}

\newcolumntype{C}{>{\centering\arraybackslash}X}
\newcolumntype{R}{>{\raggedleft\arraybackslash}X}
\newcommand{\V}{$\times$}
\newcommand{\X}{$\checkmark$}

\Crefname{figure}{Fig.}{Fig.}
\Crefname{equation}{Eq.}{Eq.}

\makeatletter
\renewcommand\NAT@sort@cites[1]{%
  \let\NAT@cite@list\@empty
  \@for\@citeb:=#1\do{\expandafter\NAT@star@cite\@citeb\@@}%
  \if@filesw
    \expandafter\write\expandafter\@auxout
      \expandafter{\expandafter\string\expandafter\citation\expandafter{\NAT@cite@list}}%
  \fi
  \@ifnum{\NAT@sort>\z@}{%
    \expandafter\NAT@sort@cites@\expandafter{\NAT@cite@list}%
  }{}%
}%
\makeatother

\begin{document}

\title{%
    SALSA: Spatial Cue-Augmented\\Log-Spectrogram Features for Polyphonic\\Sound Event Localization and Detection%
}
%
%
%

\author{Thi~Ngoc~Tho~Nguyen$^{\star}$,
        Karn~N.~Watcharasupat,~\IEEEmembership{Student~Member,~IEEE},
        Ngoc~Khanh~Nguyen,
        Douglas~L.~Jones,~\IEEEmembership{Fellow,~IEEE},
	    and Woon-Seng~Gan,~\IEEEmembership{Senior~Member,~IEEE}
\thanks{This research was supported by the Singapore Ministry of Education Academic Research Fund Tier-2, under research grant MOE2017-T2-2-060, and the Google Cloud Research Credits program with the award GCP205559654.}%
\thanks{T. N. T. Nguyen, K. N. Watcharasupat, and W.-S. Gan are with the School of Electrical and Electronic Engineering, Nanyang Technological University, 639798, Singapore (e-mail: \{nguyenth003, karn001\}@e.ntu.edu.sg, ewsgan@ntu.edu.sg). K. N. Watcharasupat further acknowledges the support from the CN Yang Scholars Programme, Nanyang Technological University, Singapore.}
\thanks{D. L. Jones is with the Department of Electrical and Computer Engineering, University of Illinois at Urbana-Champaign, IL 61801, USA (email: dl-jones@illinois.edu).}}

%
%

\markboth{IEEE/ACM Transactions on Audio, Speech, and Language Processing}%
{Nguyen \MakeLowercase{\textit{et al.}}: SALSA}
%



\maketitle

\begin{abstract}
Sound event localization and detection (SELD) consists of two subtasks,  which are sound event detection and direction-of-arrival estimation. While sound event detection mainly relies on time-frequency patterns to distinguish different sound classes, direction-of-arrival estimation uses amplitude and/or phase differences between microphones to estimate source directions. As a result, it is often difficult to jointly optimize these two subtasks. We propose a novel feature called \aname{Spatial cue-Augmented Log-SpectrogrAm} (SALSA) with exact time-frequency mapping between the signal power and the source directional cues, which is crucial for resolving overlapping sound sources. The SALSA feature consists of multichannel log-spectrograms stacked along with the normalized principal eigenvector of the spatial covariance matrix at each corresponding time-frequency bin. Depending on the microphone array format, the principal eigenvector can be normalized differently to extract amplitude and/or phase differences between the microphones. As a result, SALSA features are applicable for different microphone array formats such as first-order ambisonics (FOA) and multichannel microphone array (MIC). Experimental results on the TAU-NIGENS Spatial Sound Events 2021 dataset with directional interferences showed that SALSA features outperformed other state-of-the-art features. Specifically, the use of SALSA features in the FOA format increased the F1 score and localization recall by \SI{6}{\percent} each, compared to the multichannel log-mel spectrograms with intensity vectors. For the MIC format, using SALSA features increased F1 score and localization recall by \SI{16}{\percent} and \SI{7}{\percent}, respectively, compared to using multichannel log-mel spectrograms with generalized cross-correlation spectra.  

\end{abstract}

\begin{IEEEkeywords}
deep learning, feature extraction, microphone array, spatial cues, sound event localization and detection.
\end{IEEEkeywords}

%
\IEEEpeerreviewmaketitle

\section{Introduction}

\IEEEPARstart{S}{ound} event localization and detection (SELD) has many applications in urban sound sensing~\cite{Salamon2017DeepClassification}, wildlife monitoring~\cite{Stowell2016BirdChallenge}, surveillance~\cite{Foggia2016AudioSounds}, autonomous driving, 
and robotics~\cite{Valin2004LocalizationApproach}. SELD is an emerging research field that unifies the tasks of sound event detection (SED) and direction-of-arrival estimation (DOAE) by jointly recognizing the sound classes, and estimating the directions of arrival (DOA), the onsets, and the offsets of detected sound events~\cite{Adavanne2019SoundNetworks}. Because of a need for source localization, SELD typically requires multichannel audio inputs from a microphone array, which has several formats in current use, such as first-order ambisonics (FOA) and multichannel microphone array (MIC).  

\subsection{Existing methods}

\begin{table*}[t]
    \centering
    \caption {Comparison of the proposed method with some existing deep learning-based methods for polyphonic SELD.}  
    \label{table:seld_literatures}
    {
\setlength\tabcolsep{3pt}
\begin{tabularx}{\textwidth}{l@{ }llXll}
\toprule
    Approach & &
    Format & 
    Input Features & 
    Network Architecture &
    Output\\
\midrule 
    Adavanne\etal{} & \cite{Adavanne2019SoundNetworks} &   
        FOA/MIC & 
        Magnitude \& phase spectrograms &   
        End-to-end CRNN &   
        class-wise\\
    Cao\etal{} & \cite{Cao2019PolyphonicStrategy} &   
        FOA/MIC    & 
        Log-mel spectrograms, GCC-PHAT &   
        Two-stage CRNNs    &  
        class-wise   \\
    Nguyen\etal{} & \cite{Nguyen2020ADetection}           &   
        FOA         & 
        Log-mel spectrograms, {directional SS histograms}    &   Sequence matching CRNN   &   
        track-wise  \\
    Xue\etal{} & \cite{Xue2020SoundLearning}    &   
        MIC     & 
        Log-mel spectrograms, IV, {pair-wise phase differences}  &   
        Modified two-stage CRNNs    & 
        class-wise \\
    Cao\etal{} & \cite{Cao2021AnDetection}            &   
        FOA         & 
        Log-mel spectrograms, IV &   
        EINv2 & track-wise    \\ 
    Shimada\etal{} & \cite{Shimada2021ACCDOA:Detection}   &   
        FOA         & 
        Linear amplitude spectrograms, IPD     & 
        CRNN with D3Net   & 
        class-wise  \\
    Sato\etal{} & \cite{Sato2021AmbisonicEquivariance}   &   
        FOA         & 
        Complex spectrograms                      & 
        Invariant CRNN    &  
        class-wise \\
    Phan\etal{} & \cite{Phan2020OnLocalization}       &   
        FOA/MIC    & 
        Log-mel spectrograms, IV, GCC-PHAT  &   
        CRNN with self attention     & 
        class-wise \\
    Park\etal{} & \cite{Park2020SoundFunctions}       &   
        FOA         & 
        Log-mel spectrograms, IV, {harmonic percussive separation}  &   
        CRNN with feature pyramid   & 
        class-wise \\
    Emmanuel\etal{} & \cite{Emmanuel2021Multi-scaleDetection} &   
        FOA & 
        Constant-Q spectrograms, log-mel spectrograms, IV  &   
        Multi-scale network with MHSA   & 
        track-wise \\
    Lee\etal{} & \cite{Lee2021SoundChallenge}   &   
        FOA         & 
        Log-mel spectrograms, IV     & 
        EINv2 with cross-model attention   & 
        track-wise \\
\midrule
    (Top'19) Kapka\etal{} & \cite{Kapka2019SoundModels}     &   
        FOA         & 
        Log-mel spectrograms, IV     & 
        Ensemble of CRNNs &   
        class-wise \\
    (Top'20) Wang\etal{} & \cite{Wang2020TheChallenge}   &   
        FOA+MIC   & 
        Log-mel spectrograms, IV, GCC-PHAT  & 
        Ensemble of CRNNs \& CNN-TDNNs   &   
        class-wise \\
    (Top'21) Shimada\etal{} & \cite{Shimada2021EnsembleDetection} &   
        FOA     & 
        Linear amplitude spectrograms, IPD, cosIPD, sinIPD     & 
        Ensemble of CRNNs \& EINv2  & 
        class-wise \\
\midrule 
    Proposed method    & &
        FOA/MIC  & 
        SALSA: Log-linear spectrograms \& normalized eigenvectors & 
        End-to-end CRNN &     
        class-wise \\ 
\bottomrule 
\end{tabularx}
} 
    
    \begin{justify}
        IV and \gccphat features follow the frequency scale (linear, mel, constant-Q) of the spectrograms. TDNN stands for time delay neural networks. IPD stands for interchannel phase differences. Top'YY denotes the top ranked systems for the respective DCASE SELD Challenges.
    \end{justify}
\end{table*}

Over the past few years, there have been many major developments for SELD in the areas of data augmentation, feature engineering, model architectures, and output formats. In 2015, an early monophonic SELD work by Hirvonen~\cite{Hirvonen2015ClassificationNetworks} formulated SELD as a classification task. In 2018, Adavanne\etalcite{Adavanne2019SoundNetworks} pioneered a seminal polyphonic SELD work that used an end-to-end convolutional recurrent neural network (CRNN), \aname{SELDnet}, to jointly detect sound events and estimate the corresponding DOAs. In 2019, SELD task was introduced in the Challenge on Detection and Classification of Acoustic Scenes and Events (DCASE). Cao\etalcite{Cao2019PolyphonicStrategy} proposed a two-stage strategy by training separate SED and DOAE models.
Mazzon\etalcite{Mazzon2019FirstEstimation} proposed a spatial augmentation method by swapping channels of FOA format. 
Nguyen\etalcite{Nguyen2020ADetection, Nguyen2021ANetwork} explored a hybrid approach called a \aname{Sequence Matching Network} (SMN) that matched the SED and DOAE output sequences using a bidirectional gated recurrent unit (BiGRU).

In 2020, moving sound sources were introduced in the DCASE SELD Challenge. Cao\etalcite{Cao2020Event-independentDetection} proposed \aname{Event Independent Network} (EIN) that used soft parameter sharing between the SED and DOAE encoder branches and output track-wise predictions. An improved version of this network, EINv2, replaced the biGRUs with multi-head self-attention (MHSA)~\cite{Cao2021AnDetection}. Sato\etalcite{Sato2021AmbisonicEquivariance} designed a CRNN that is invariant to rotation, scale, and time translation for FOA signals. Phan\etalcite{Phan2020OnLocalization} formulated SELD as regression problems for both SED and DOAE to improve training convergence. 
Wang\etalcite{Wang2021ADetection} focused on several data augmentation methods to overcome the data sparsity problem in SELD. Shimada\etalcite{Shimada2021ACCDOA:Detection} unified SED and DOAE losses into one regression loss using a representation technique called \aname{Activity-Coupled Cartesian Direction of Arrival} (ACCDOA).
In 2021, unknown interferences were introduced in the DCASE SELD challenge. Lee\etalcite{Lee2021SoundChallenge} enhanced EINv2 by adding cross-modal attention between the SED and DOAE branches. 
\Cref{table:seld_literatures} summarizes some notable and state-of-the-art deep learning methods for SELD.

\subsection{Input features for SELD}

In this paper, we focus on input features for SELD. When SELDnet was first introduced, it was trained on multichannel magnitude and phase spectrograms~\cite{Adavanne2019SoundNetworks}. Subsequently, different features, such as multichannel log-spectrograms and intensity vector (IV) for the FOA format, and generalized cross-correlation with phase transform (\gccphat) for the MIC format in the mel scale were shown to be more effective for SELD~\cite{Politis2020ADetection, Politis2021, Cao2019PolyphonicStrategy, Cao2021AnDetection, Shimada2021ACCDOA:Detection, Wang2021ADetection, Wang2020TheChallenge, Shimada2021EnsembleDetection}. 

Due to the smaller dimension size and stronger emphasis on the lower frequency bands, where signal contents are mostly populated, the mel frequency scale has been used more frequently than the linear frequency scale for SELD. However, combining the IV or \gccphat features with the mel spectrograms is not trivial and the implicit DOA information stored in the former features are often compromised. In practice, the IVs are also passed through the mel filters which merge DOA cues in different narrow bands into one mel band, making it more difficult to resolve different DOAs in multi-source scenarios. Likewise, in order to stack the \gccphat with the mel spectrograms, longer time-lags on the \gccphat have to be truncated. Since the linear scale has the advantage of preserving the directional information at each frequency band, several works have attempted to use spectrogram, inter-channel phase differences (IPD), and IVs in linear scale~\cite{Shimada2021ACCDOA:Detection} or the constant-Q scale~\cite{Emmanuel2021Multi-scaleDetection}. However, there is lack of experimental results that directly compare these features over different scales. 

Referring to \Cref{table:seld_literatures}, more SELD algorithms have been developed for the FOA format compared to the MIC format, even though the MIC format is more common in practice. The baselines for three DCASE SELD challenges so far have indicated that using FOA inputs performs slightly better than that with MIC inputs~\cite{Adavanne2019ADetection, Politis2020ADetection, Politis2021}.  
In addition, it is more straightforward to stack IVs with the spectrograms in the FOA format compared to stacking \gccphat with spectrograms. 
When IVs are stacked with spectrograms, there is a direct frequency correspondence between the IVs and the spectrograms. This frequency correspondence is crucial for networks to associate the sound classes and the DOAs of multiple sound events, where signals of different sound sources are often distributed differently along the frequency dimension.
On the other hand, the time-lag dimension of the \gccphat features does not have a local linear one-to-one mapping with the frequency dimension of the spectrograms. As a result, all of the DOA information is aggregated at the frame level, making it difficult to assign correct DOAs to different sound events. Furthermore, when there are multiple sound sources, \gccphat features are known to be noisy, and the directional cues at overlapping TF bins of IVs are merged.
In order to solve SELD more effectively in noisy, reverberant, and multi-source scenarios, a better feature is needed for both audio formats, but especially for the MIC format where feature engineering has largely been lacking compared to the FOA format.

\subsection{Our Contributions}

We propose a novel feature for SELD called Spatial Cue-Augmented Log-Spectrogram (SALSA) with exact spectrotemporal mapping between the signal power and the source DOA for both FOA and MIC formats. The feature consists of multichannel log-magnitude linear-frequency spectrograms stacked with a normalized version of the principal eigenvector of the spatial covariance matrix at each TF bin on the spectrograms. The principal eigenvector is normalized such that it represents the inter-channel intensity difference (IID) for the FOA format, and/or inter-channel phase difference (IPD) for the MIC format. 

To further improve the performance, only eigenvectors from approximately single-source TF bins are included in the features since these directional cues are less noisy. A TF bin is considered a single-source bin when it contains energy mostly from only one source~\cite{Nguyen2014RobustSources, Nguyen2020RobustNetwork}. We evaluated the effectiveness of the proposed feature on both the FOA and the MIC formats using the TAU-NIGENS Spatial Sound Events (TNSSE) 2021 dataset used in DCASE 2021 SELD Challenge. Experimental results showed that the SALSA feature outperformed, for the FOA format, both mel- and linear-frequency log-magnitude spectrograms with IV, and for the MIC format, the log-magnitude spectrogram with \gccphat.

In addition, SALSA features bridged the performance gap between the FOA and the MIC formats, and achieved the state-of-the-art performance for a single (non-ensemble) model on the TNSSE 2021 development dataset for both formats. Similarly, when evaluated on the TNSSE 2020 dataset, SALSA also achieved the top performance for a single model for both formats on both the development and the evaluation datasets. Our ensemble model trained on an early version of SALSA features ranked second in the team category of the DCASE 2021 SELD challenge~\cite{Nguyen2021DCASEDetection}.   

Our paper offers several contributions, as follows:
\begin{enumerate}
    \item a novel and effective feature for SELD that works for both FOA and MIC formats,
    \item an improvement to the proposed feature by utilizing signal processing-based methods to select single-source TF bins,
    \item a comprehensive analysis of feature importance of each components in SALSA for SELD, and,
    \item an extensive ablation study of different data augmentation methods for the newly proposed SALSA feature, as well as for the log-magnitude spectrograms with IV and \gccphat in both linear- and mel-frequency scales.
\end{enumerate}

The rest of the paper is organized as follows. \Cref{sec:salsa} presents the proposed SALSA features for both the FOA and the MIC formats. \Cref{sec:common_features} briefly describes common SELD features used as benchmarks. \Cref{sec:net} presents the network architecture employed in all of the experiments. \Cref{sec:exp} elaborates the experimental settings. \Cref{sec:results} presents the experimental results and discussion with extensive ablation study. Finally, we conclude the paper in \Cref{sec:conclusion}. The source code for reproducing our work can be found at \href{https://github.com/thomeou/SALSA}{https://github.com/thomeou/SALSA}.
\section{Spatial Cue-Augmented Log-Spectrogram Features for SELD}
\label{sec:salsa}

The proposed SALSA features consist of two major components: multichannel log-linear spectrograms and normalized principal eigenvectors. For the rest of this paper, spectrograms refer to multichannel spectrograms unless otherwise stated. 

\subsection{Signal Model}
\label{sec:signal_model}

Let $M$ be the number of microphones and $L$ be the number of sound sources. 
The short-time Fourier transform (STFT) signal observed by an $M$-channel microphone array of arbitrary geometry in the TF domain is given by 
\begin{equation}
    \mathbf{X}(t,f)=\sum_{i=1}^{L} S_i(t,f)\mathbf{H}(f,\phi_i,\theta_i) + \mathbf{V}(t,f) \in\mathbb{C}^M,
    \label{eq:stft_signal_model}
\end{equation}
where $t$ and $f$ are time and frequency indices, respectively; $S_i$ is the $i$th source signal; $\mathbf{H}(f,\phi_i, \theta_i)$ is the frequency-domain steering vector corresponding to the DOA ($\phi_i,\theta_i$) of the $i$th source, where $\phi_i$ and $\theta_i$ are the azimuth and elevation angles, respectively; and $\mathbf{V}$ is the noise vector. For moving sources, $\phi_i=\phi_i(t)$ and $\theta_i=\theta_i(t)$ are functions of time. For brevity, the time variable is omitted in $\phi_i$ and $\theta_i$ for some equations. Note that \Cref{eq:stft_signal_model} is applicable for TF bins that have relatively low reverberation, which can be absorbed into the $\mathbf{V}$ term. TF bins with relative high direct-to-reverberant energy ratios would be preferably excluded from the estimation in this work. 

\subsection{Multichannel log-linear spectrograms}

The log-linear spectrograms are computed from the complex spectrograms $\mathbf{X}(t,f)$ by
\begin{equation}
    \textsc{LinSpec}(t,f) = \log\left(\left\lvert\mathbf{X}(t,f)\right\rvert^2\right) \in\mathbb{R}^{M\times T\times F},
    \label{eq:linspec}
\end{equation}
where $\lvert\cdot\rvert$, is the elementwise complex modulus, $T$ is the number of time frames and $F$ is the number of frequency bins.

\subsection{Normalized principal eigenvectors}

Assuming the signal and noise are zero-mean and uncorrelated, the true covariance matrix, $\mathbf{R}(t,f) \in\mathbb{C}^{M\times M}$, is a linear combination of rank-one outer products of the steering vectors weighted by signal powers $\sigma_i^{2}(t,f)$ of the $i$th source at the TF bin $(t,f)$, that is,
\begin{align}
    & \mathbf{R}(t,f) = \E[\mathbf{X}(t,f)\mathbf{X}^{\H}(t,f)]\label{eq:ex_covmat}\\
    & = \sum_{i=1}^{L} \sigma_i^{2}(t,f)\mathbf{H}(f,\phi_i, \theta_i) \mathbf{H}^{\H}(f,\phi_i, \theta_i) + \mathbf{R}_\text{n}(t,f),\label{eq:true_covmat}
\end{align}
where $\mathbf{R}_\text{n}(t,f)$ is the noise covariance matrix, and $(\cdot)^{\mathbf{\H}}$ denotes the Hermitian transpose. Note that although the reverberation is also absorbed into the noise vector, the uncorrelated noise assumption can generally hold if the reverberation level is sufficiently low.

In practice, under the assumption that the sources are slow-moving within a small time window, \Cref{eq:ex_covmat} can be approximated using
\begin{equation}
    \mathbf{\hat{R}}(t,f)= \frac{1}{2T_\text{r}+1}\sum_{\tau=-T_\text{r}}^{T_\text{r}}\mathbf{X}(t+\tau,f)\mathbf{X}^{\H}(t+\tau,f)
\end{equation}
where $2T_\text{r}+1$ is the window size. In this work, we use $T_\text{r}=3$.

\Cref{eq:true_covmat} shows that at single-source TF bins, where only one sound source is dominant over other sources and reverberation, the theoretical steering vector $\mathbf{H}(f,\phi, \theta)$ can be approximated by the principal eigenvector $\mathbf{U}(t, f)$ of the covariance matrix, a technique previously utilized in multichannel speech separation~\cite{Asano2007DetectionArray}, as well as in our previous works \cite{Nguyen2014RobustSources, Nguyen2020RobustNetwork}.
Therefore, we can reliably extract directional cues from these principal eigenvectors at these bins. 
For TF bins which are not single-source, the values of the directional cues can be set to a predefined default value such as zero. In the next sections, we elaborate on how to normalize the principal eigenvectors to extract directional cues, which are encoded in the IID and IPD for FOA arrays and far-field microphone arrays, respectively.

\subsubsection{Eigenvector-based intensity vector for FOA arrays}
\label{subsec:eiv}

FOA arrays have four channels and the directional cues are encoded in the IID. A typical steering vector for an FOA array can be defined by
\begin{equation}
    \mbf{H}^\text{FOA}(t,\phi, \theta)
    = \begin{bmatrix}
    H_{\text{W}}(t,\phi, \theta) \\
    H_{\text{X}}(t,\phi, \theta) \\
    H_{\text{Y}}(t,\phi, \theta) \\
    H_{\text{Z}}(t,\phi, \theta) 
    \end{bmatrix} = \begin{bmatrix}
    1 \\
    \cos(\phi) \cos(\theta) \\
    \sin(\phi) \cos(\theta) \\
    \sin(\theta)
    \end{bmatrix} \in \mathbb{R}^{4},
\label{eq:foa_steervec}
\end{equation} \nobreak
where $\phi=\phi(t)$ and $\theta=\theta(t)$ are the time-dependent azimuth and elevation angles of a sound source with respect to the array, respectively.

We can compute an eigenvector-based intensity vector (EIV) to approximate $[H_{\text{X}}, H_{\text{Y}}, H_{\text{Z}}]^\T$ from the principle eigenvector ${\mathbf{U}=\mathbf{U}(t,f)}$ as follows. First, we normalize $\mathbf{U}$ by its first element, which corresponds to the omni-directional channel, then discard the first element to obtain $\mathbf{\bar{U}}$. Afterwards, we take the real part of $\mathbf{\bar{U}}$ and normalize it to obtain unit-norm EIV $\mathbf{\widetilde{U}} = \Re(\mathbf{\bar{U}})/\|\Re(\mathbf{\bar{U}})\|$. SALSA features for the FOA format are formed by stacking the four-channel spectrograms with the three-channel EIV $\mathbf{\widetilde{U}}$. 

\begin{figure}[tb]
    \centering
    \includegraphics[width=\linewidth,bb=0 0 729 702]{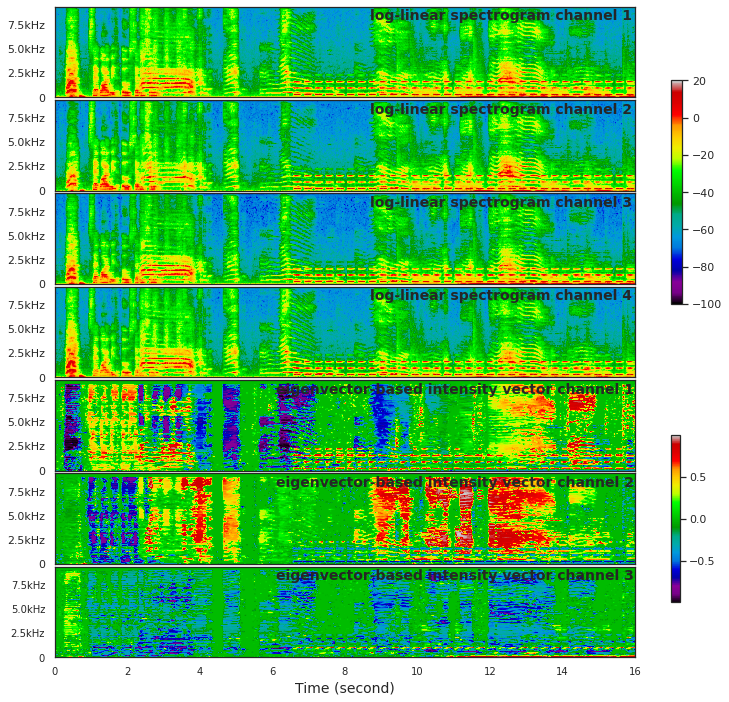}
    \vspace{-0.5cm}
    \caption{SALSA features of a 16-second audio segment of FOA format in a multi-source scenario. 
    The vertical axis represents frequency in kHz. In the spectrogram channels, the colormap represents the signal log-magnitude in each TF bin. In the EIV channels, the colormap represents the values of the computed EIV features.}
    \label{fig:foa_salsa}
\end{figure}

\Cref{fig:foa_salsa} illustrates SALSA features of a \num{16}-second audio segment in multi-source cases for an FOA array with an EIV cutoff frequency of \SI{9}{\kilo\hertz}. The three EIV channels are visually discriminant for different sources originating from different directions. The green areas in the EIV channels correspond to zeroed-out TF bins. Moreover, due to the spectrotemporal alignment properties of SALSA, it can be observed that the TF patterns of the sources in the spectrogram channels, and the patterns of the corresponding directional cues share similar activation patterns, facilitating multichannel feature extraction that is also spectrotemporally meaningful when used as an input to convolutional layers.

\subsubsection{Eigenvector-based phase vector for microphone arrays}
\label{subsec:epv}

\begin{figure}[tb]
    \centering
    \includegraphics[width=\linewidth,bb=0 0 729 702]{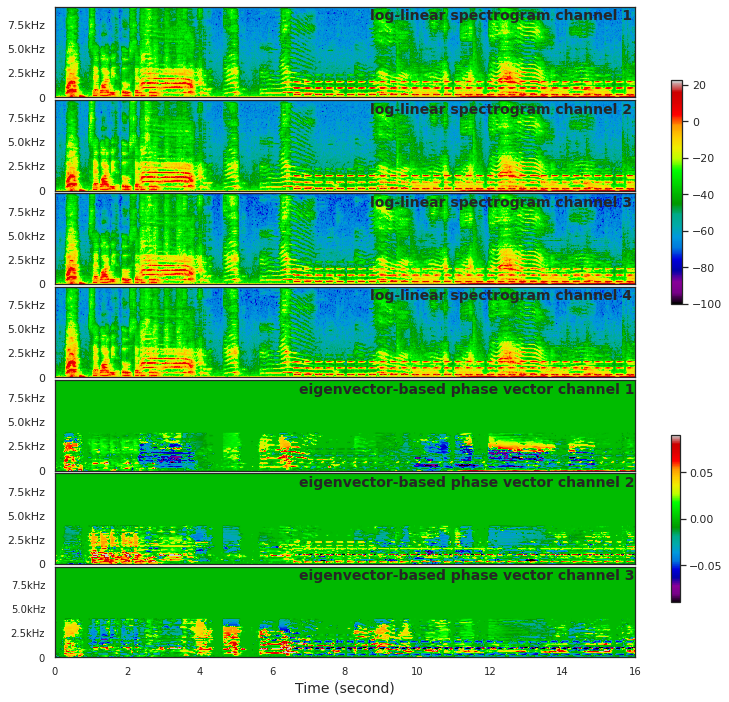}
    \vspace{-0.5cm}
    \caption{SALSA features of a 16-second audio segment of a four-channel microphone array (MIC format) in a multi-source scenario. The vertical axis represents frequency in kHz. In the spectrogram channels, the colormap represents the signal log-magnitude in each TF bin. In the EPV channels, the colormap represents the values of the computed EPV features.}
    \label{fig:mic_salsa}
\end{figure} 

For a far-field microphone array, the directional cues are encoded in the IPD. The steering vector of an $M$-channel far-field array of an arbitrary geometry can be modelled by $\mbf{H}^\text{MIC}(t, f, \phi, \theta)\in \mathbb{C}^{M}$, whose elements are given by
\begin{equation}
    H_m^\text{MIC}(t, f, \phi, \theta)
    = \exp\left(-\jmath 2\pi f d_{1m}(\phi(t), \theta(t)) / c\right),
\label{eq:mic_steervec}
\end{equation} \nobreak
where $\jmath$ is the imaginary unit, $c \approx \SI{343}{\meter\per\second}$ is the speed of sound; $d_{1m}(\phi(t), \theta(t))$ is the distance of arrival, in metres, travelled by a sound source, between the $m$th microphone and the reference ($m=1$) microphone. In theory, the distance of arrival is given by
\begin{equation}
    d_{1m}(\phi(t), \theta(t)) = (\bm{\zeta}_1 - \bm{\zeta}_m)^\T 
    \begin{bmatrix} 
        \cos(\phi(t)) \cos(\theta(t)) \\ 
        \sin(\phi(t)) \cos(\theta(t)) \\ 
        \sin(\theta(t))
    \end{bmatrix} \in \mathbb{R},
    \label{eq:dist_of_arrival}
\end{equation}
where $\bm{\zeta}_1$ and $\bm{\zeta}_m$ are the Cartesian coordinates of the reference and the $m$th microphones, respectively. $\tau_{1m}(\phi(t), \theta(t)) = d_{1m}(\phi(t), \theta(t)) / c$ is the time difference of arrival (TDOA), travelled by the sound source, between the $m$th and the reference microphones. 

The directional cues of a far-field microphone array can be presented in several forms such as the relative distance of arrival (RDOA) and TDOA. In this study, we choose to extract the directional cues in the form of RDOA. One advantage of RDOA is that we do not need to know the exact coordinates of the individual microphones, since spatial information of the microphones are already implicitly encoded in the RDOA. We can compute an eigenvector-based phase vector (EPV) to approximate $[d_{12}, \ldots, d_{1M}]^T$ from the principle eigenvector $\mathbf{U}$ as follows. First we normalize $\mathbf{U}$ by its first element, which is chosen arbitrarily as the reference microphone, then discard the first element to obtain $\mathbf{\bar{U}}$. After that, we take the phase of $\mathbf{\bar{U}}$ and normalize it by $-2\pi f/c$ to obtain the EPV $\mathbf{\widetilde{U}} =- c\angle\mathbf{\bar{U}}/(2 \pi f)$. The SALSA features for {far-field microphone arrays} are formed by stacking the $M$-channel spectrograms with the $(M-1)$-channel EPV. To avoid spatial aliasing, the values of $\mathbf{\widetilde{U}}$ are set to zero for all TF bins above aliasing frequency. 

\Cref{fig:mic_salsa} illustrates the SALSA feature of a \num{16}-second audio segment in multi-source cases for a four-channel microphone array with an EPV cutoff frequency of \SI{4}{\kilo\hertz}. Similar to the FOA counterpart, the three EPV channels are visually discriminant for different sources originating from different directions. The directional cues in the EPV channels also similarly display patterns corresponding to the sources. The green areas in the EPV channels corresponds to zeroed-out TF bins that are not {single-source} or above aliasing frequency. 

The proposed method to extract spatial cues can also be extended to near-field and baffled microphone arrays, where directional cues are encoded in both IID and IPD. For those arrays, we can approximate their array response model using the far-field model, or we can compute both EIV and EPV as shown in \Cref{subsec:eiv} and \Cref{subsec:epv}, respectively.

\subsection{Single-source time-frequency bin selection}
\label{subsec:ss_tf_selection}

\begin{figure}[tb]
    \centering
    \includegraphics[width=0.8\columnwidth,bb=0 0 439 299]{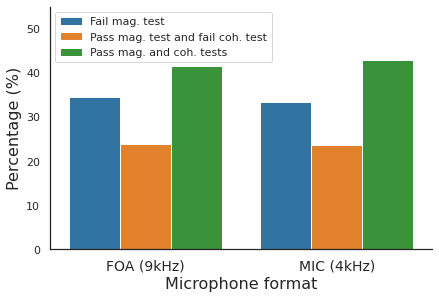}
    \vspace{-0.3cm}
    \caption{Distributions of TF bins that {(a)} fail magnitude test, {(b)} pass magnitude test but fail coherence test, and {(c)} pass both tests, for the FOA and MIC formats. The distributions shown are independent of the true number of sources.}
    \label{fig:tf_bin_dist}
\end{figure} 

The selection of single-source TF bins have been shown to be effective for DOAE in noisy, reverberant and multi-source cases~\cite{Nguyen2014RobustSources, Nguyen2020ADetection, Nguyen2020RobustNetwork}. There are several methods to select single-source TF bins~\cite{Mohan2008LocalizationTest, Pavlidi2013Real-TimeArray, Nguyen2014RobustSources}. In this paper, we apply two tests to select single-source TF bins, namely, the magnitude and coherence tests. The magnitude test aims to select only TF bins that contain signal from foreground sound sources~\cite{Nguyen2014RobustSources}. A TF bin passes the magnitude test if its signal-to-noise ratio (SNR) with respect to the adaptive noise floor $\eta[t, f]$ is above a threshold $\alpha_\text{SNR}$~\cite{Nguyen2014RobustSources}. In practice, the magnitude test indicator is given by
\begin{equation}
    \textsc{MagTest}[t, f] = \mathbb{I}\left[
        \widetilde{X}_1[t,f] > \alpha_\text{SNR} \cdot \eta[t, f]
    \right] \in \{0, 1\},
\end{equation}
where $\mathbb{I}\left[\cdot\right]$ is the Iverson bracket, and $\widetilde{X}_1[t,f]$ is a running root-mean-square of the magnitude of $X_1[t,f]$ over a 3-frame window. We set the threshold $\alpha_\text{SNR}=1.5$ based on past experiments across several DOAE \cite{Nguyen2014RobustSources, Nguyen2020RobustNetwork, Zhao2015RobustReduction}, and SELD \cite{Nguyen2020ADetection, Nguyen2020EnsembleTracking, Nguyen2019DCASEDetection} tasks. We used a fast and simple method to estimate the frequency-wise noise floor $\eta[t, f]$ as follows~\cite{Nguyen2014RobustSources}. The noise floor is initialized using the first few audio frames, which are assumed to contain only noise. After that, the noise floor is slightly increased or decreased if the magnitude of $X_1[t,f]$ is above or below the previous noise floor, respectively. The noise floor can also be computed using other estimators such as the one proposed in \cite{Gerkmann2012UnbiasedDelay}.

The coherence test aims to find TF bins that contain signal from mostly one source~\cite{Mohan2008LocalizationTest}. Specifically, consider an eigendecomposition, which in practice can be equivalently performed by the more numerically stable SVD,
\begin{equation}
    \mathbf{R}(t,f) = \mathbf{U}(t,f)\mathbf{\Sigma}(t,f)\mathbf{U}^\H(t,f)
\end{equation}
where $\mathbf{\Sigma}(t,f)=\operatorname{diag}\left([\sigma_1(t,f), \sigma_2(t,f),\cdots,\sigma_M(t,f)]\right)$, and $\sigma_1(t,f)\ge\sigma_2(t,f)\ge\dots\ge\sigma_M(t,f)$. The direct-to-reverberant ratio (DRR) is given by
\begin{equation}
    \rho(t,f) = \dfrac{\sigma_1(t,f)}{\sigma_2(t,f)}.
\end{equation}
Since the DRR can be interpreted as the relative strength between the paths of signals arriving at the microphone array, the DRR can be also interpreted as a measure of source dominance. When the DRR is low, it is likely that there are either multiple dominant sources at the TF bin, or the reverberation is high even if there is only one source.
A TF bin passes the coherence test if its DRR is above a coherence threshold $\beta_\text{DRR}$~\cite{Rafaely2017SpeakerStatistics}. In this work, we used $\beta_\text{DRR}=5$ which obtained the best performance on the validation based on a grid search. 

\Cref{fig:tf_bin_dist} shows the distribution of TF bins that fail magnitude test, pass magnitude test but fail coherence test, and pass both tests for the FOA and MIC formats from the TNSSE 2021 development dataset~\cite{Politis2021}. The lower cutoff frequency for both formats is \SI{50}{\hertz} while the upper cutoff frequency for the FOA and MIC formats are \SI{9}{\kilo\hertz} and \SI{4}{\kilo\hertz}, respectively. For both formats, around \SI{40}{\percent} of TF bins in the passband pass both tests. The two tests significantly reduce the number of EIVs or EPVs to be computed. 
\section{Common input features for SELD}
\label{sec:common_features}

\begin{table}[t]
    \centering
    \caption {Feature names and descriptions}  
    \label{tab:features}
    \noindent\begin{tabularx}{\columnwidth}{llXr}
    \toprule
    Name & Format &  Components & \# channels \\ 
    \midrule
    \textsc{MelSpecIV}       & FOA   & \textsc{MelSpec} + IV & \num{7} \\
    \textsc{LinSpecIV}       & FOA   & \textsc{LinSpec} + IV  & \num{7}\\
    \textsc{MelSpecGCC}       & MIC   & \textsc{MelSpec} + GCC-PHAT & \num{10}\\
    \textsc{LinSpecGCC}       & MIC   & \textsc{LinSpec} + GCC-PHAT  & \num{10}\\
    \midrule
    SALSA           & FOA   & \textsc{LinSpec} + EIV & \num{7} \\
    SALSA           & MIC   & \textsc{LinSpec} + EPV & \num{7} \\
    \bottomrule
    \end{tabularx}
    \begin{justify}
        The number of channels are calculated based on four-channel inputs.
    \end{justify}
\end{table}

We compare the proposed SALSA features with log-spectrograms and IV for the FOA format, and log-spectrograms and \gccphat for the MIC format in both mel- and linear-frequency scales, of which the mel-scale features are the more popular for SELD. The log-mel spectrograms are computed from the complex spectrograms $\mathbf{X}$ by
\begin{equation}
    \textsc{MelSpec}(t,k) = \log\left(\left\lvert\mathbf{X}(t,f)\right\rvert^2 \cdot \mathbf{W}_\text{mel}(f,k)\right),
    \label{eq:melspec}
\end{equation}
where $k$ is the mel index, and $\mathbf{W}_\text{mel}$ is the mel filter. 

\subsection{Log-spectrograms and IV for FOA format}

The four channels of the FOA format consist of the \mbox{omni-,} X-, Y-, and Z-directional components. The IV expresses intensity differences of the X, Y, and Z components with respect to the omni-directional component, and thus carries the DOA cues~\cite{Zhao2014UnderdeterminedSensor, Cao2019Two-StageCross-Correlation}. The active IV is computed in the TF domain by
\begin{align}
    \mathbf{\Lambda}(t,f) = -\frac{1}{\epsilon_0 c}\Re\left[ 
    X_\text{W}^{*}(t,f) 
    \begin{pmatrix}
        X_\text{X}(t,f)\\
        X_\text{Y}(t,f)\\
        X_\text{Z}(t,f)
    \end{pmatrix}
    \right],
    \label{eq:iv}
\end{align}
where $\epsilon_0$ is the sound density~\cite{Cao2020Event-independentDetection}. Physically, the active IV corresponds to the flow of acoustic energy thus the directional cues of the location(s) of sound source(s) can be extracted~\cite{Delikaris-Manias2017DOAVectors}. The IV features are then normalized~\cite{Cao2020Event-independentDetection} to have unit norm via $\mathbf{\bar{\Lambda}}(f,t) = \mathbf{\Lambda}(f,t) / \| \mathbf{\Lambda}(f,t) \|$. In order to combine IVs and the multichannel log-mel spectrograms, the IVs are passed through the same set of mel filters $\mathbf{W}_\text{mel}$ used to compute the log-mel spectrograms; we refer to this feature as \textsc{MelSpecIV}. Linear-scale IV can also be stacked with log-linear spectrograms, referred to as \textsc{LinSpecIV}. The dimensions of \textsc{MelSpecIV} and \textsc{LinSpecIV} are $7 \times T \times K$ and $7 \times T \times F$, respectively, where $K$ is the number of mel filters.

\subsection{Log-spectrograms and \gccphat for MIC format}

\gccphat is computed for each audio frame for each of the microphone pairs $(i, j)$ by~\cite{Cao2019PolyphonicStrategy}
\begin{equation}
    \text{GCC-PHAT}_{i,j}(t, \tau) = \mathcal{F}^{-1}_{f\rightarrow \tau}\left[
        \frac{X_i(t,f) X_j^{\H}(t,f)}{\|X_i(t,f) X^{\H}_j(t,f)\|}\right],
    \label{eq:gcc-phat}
\end{equation}
where $\tau$ is the time lag, $\mathcal{F}^{-1}$ is the inverse Fourier transform. The maximum time lag of the \gccphat spectrum is $f_\text{s} d_\text{max}/c$, where $f_\text{s}$ is the sampling rate, and $d_\text{max}$ is the largest distance between two microphones. When the \gccphat features are stacked with mel- or linear-scale spectrograms, the ranges of time lags to be included in the \gccphat spectrum are $(-K/2, K/2]$ or $(-F/2, F/2]$, respectively. We refer to these two features as \textsc{MelSpecGCC} and \textsc{LinSpecGCC}, respectively. The dimensions of the \textsc{MelSpecGCC} and \textsc{LinSpecGCC} feature are $(M + M(M-1)/2) \times T \times K$ and ${(M+M(M-1)/2) \times T \times F}$, respectively. Table~\ref{tab:features} summarizes all features of interest in this work.

\section{Network Architecture and Pipeline}
\label{sec:net}

\begin{figure}[t]
\centering
\includegraphics[width=0.9\columnwidth]{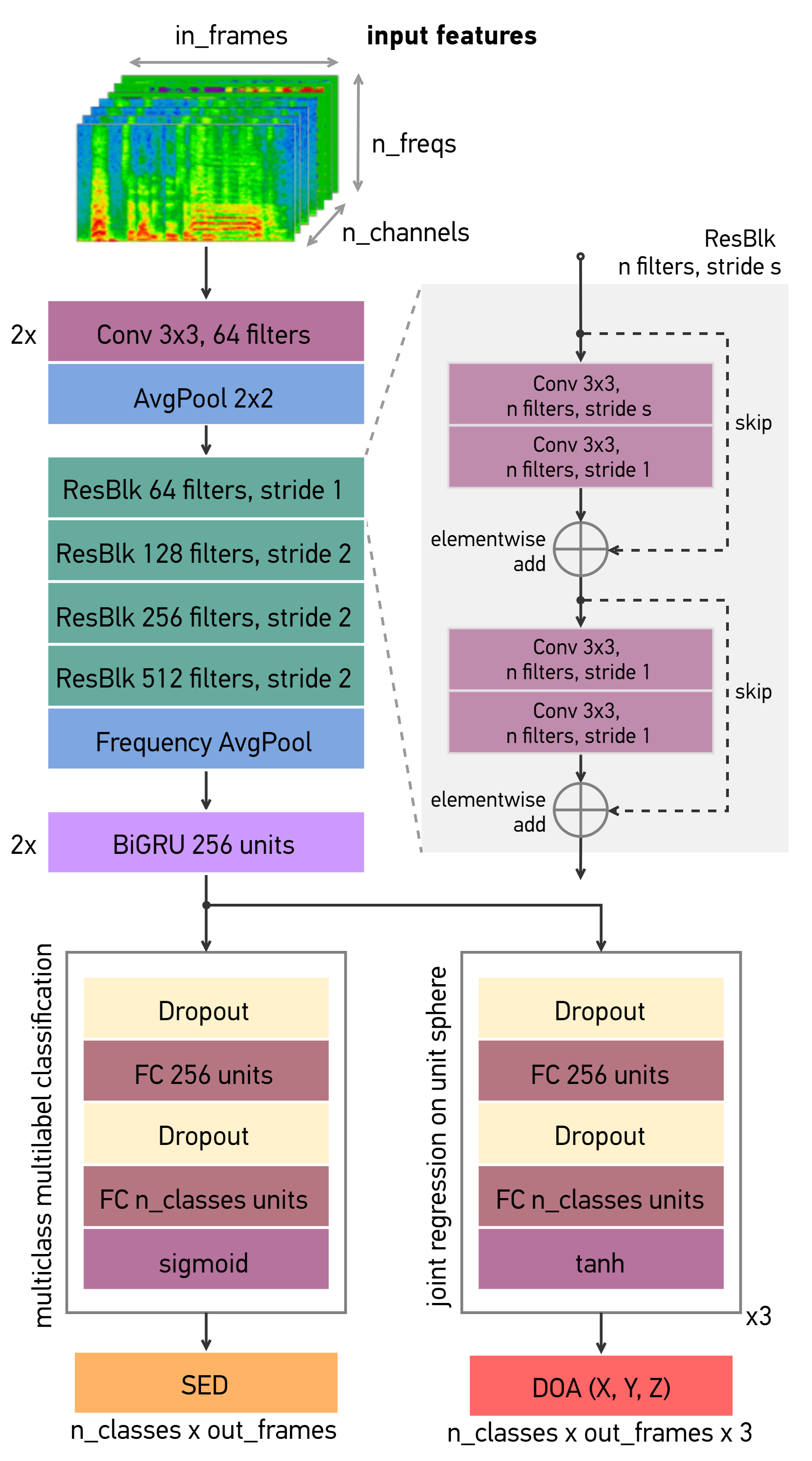}
\caption{Block diagram of the SELD network, which is a CRNN. This network can be adapted for different input features such as SALSA, \textsc{MelSpecIV}, \textsc{MelSpecGCC}, etc. by changing the number of input channels in the first convolutional layer of the network. }
\label{fig:seldnet}
\end{figure}

Figure~\ref{fig:seldnet} shows the SELD network architecture that is used for all the experiments in this paper. Since the CRNN structure is arguably the most commonly used architecture in SELD \cite{Adavanne2019SoundNetworks, Cao2019PolyphonicStrategy, Nguyen2020ADetection, Xue2020SoundLearning, Shimada2021ACCDOA:Detection, Sato2021AmbisonicEquivariance, Phan2020OnLocalization, Park2020SoundFunctions, Emmanuel2021Multi-scaleDetection, Kapka2019SoundModels, Wang2020TheChallenge, Shimada2021EnsembleDetection, Nguyen2021DCASEDetection}, we constructed the network as a CRNN, consisting of a CNN based on the PANN ResNet22 model for audio tagging~\cite{Kong2020PANNs:Recognition}, a two-layer BiGRU, and fully connected (FC) output layers. We opted to use a CNN backbone based on the PANNs~\cite{Kong2020PANNs:Recognition} given its common usage across many audio-related applications.

The network can be adapted for different input features in Table~\ref{tab:features} by setting the number of input channels in the first convolutional layer to that of the input features. During inference, sound classes whose probabilities are above the SED threshold are considered active classes. The DOAs corresponding to these classes are selected accordingly.

\subsection{Loss function}

We use the class-wise output format for SELD, in which the SED is formulated as a multilabel multiclass classification and the DOAE as a three-dimensional Cartesian regression. The loss function used is given by 
\begin{equation}
\begin{split}
    \mathcal{L}(\hat{\mbf{Y}}, \mbf{Y}) &= 
         \lambda \mathcal{L}_\textsc{bce}(\hat{\mbf{Y}}_{\textsc{sed}}, \mbf{Y}_{\textsc{sed}})\\&\quad+ \gamma \mathbbm{1}_\text{active} \mathcal{L}_\textsc{mse}(\hat{\mbf{Y}}_{\textsc{doa}}, \mbf{Y}_{\textsc{doa}}),
\end{split}
    \label{eq:loss}    
\end{equation}
where $T_\text{o}$ is the number of output frames; and $N$ is the number of target sound classes; $\hat{\mbf{Y}}, \mbf{Y} \in\mathbb{R}^{T_\text{o}\times N\times 4}$ are the SELD prediction and target tensors, respectively; $\hat{\mbf{Y}}_{\text{SED}}, \mbf{Y}_{\text{SED}} \in \mathbb{R}^{T_\text{o}\times N}$ are the SED prediction and target tensors, respectively; $\hat{\mbf{Y}}_{\text{DOA}}, \mbf{Y}_{\text{DOA}} \in \mathbb{R}^{T_\text{o}\times N\times 3}$ are the DOA prediction and target tensors, respectively. The DOA loss is only computed for the active classes in each frame.  

\subsection{Feature normalization}

The four features \textsc{MelSpecIV}, \textsc{LinSpecIV}, \textsc{MelSpecGCC}, and \textsc{LinSpecGCC} are globally normalized for zero mean and unit standard deviation vectors per channel~\cite{Adavanne2021DCASEInterference}. For the SALSA features, only the spectrogram channels are similarly normalized.

\subsection{Data augmentation}

To tackle the problem of small datasets in SELD, we investigate the effectiveness of three data augmentation techniques for all features listed in Table~\ref{tab:features}: channel swapping (CS)~\cite{Mazzon2019FirstEstimation, Wang2021ADetection}, random cutout (RC)~\cite{Zhong2020RandomAugmentation, Park2019SpecAugment:Recognition}, and frequency shifting (FS). All the three augmentation techniques can be performed in the STFT domain on the fly during training. Only channel swapping changes the ground truth, while random cutout and frequency shifting do not alter the ground truth. Each training sample has an independent \SI{50}{\percent} chance to be augmented by each of the three techniques.  

In channel swapping, there are \num{16} and \num{8} ways to swap channels for the FOA~\cite{Mazzon2019FirstEstimation} and MIC~\cite{Wang2021ADetection} formats, respectively. The IV, GCC-PHAT, EIV, EPV, and target labels are altered accordingly when channels are swapped. channel swapping augmentation technique greatly increases the variation of DOAs in the dataset. 

In random cutout, we either apply random cutout~\cite{Zhong2020RandomAugmentation} or TF masking via SpecAugment~\cite{Park2019SpecAugment:Recognition} on all the channels of the input features. Random cutout produces a rectangular mask on the spectrograms while SpecAugment produces a cross-shaped mask. For the \textsc{LinSpec} and \textsc{MelSpec} channels, the value of the mask is set to a random value within these channels' value range. For the IV, GCC-PHAT, EIV and EPV channels, the value of the mask is set to zero. All the channels share the same mask. The random cutout technique aims to improve network redundancy. 

We also introduce frequency shifting as a new data augmentation for SELD. frequency shifting in  the frequency domain is similar to pitch shift in the time domain~\cite{Salamon2017DeepClassification}. We randomly shift all the channels input features up or down along the frequency dimension by up to \num{10} bands. For \textsc{MelSpecGCC} and \textsc{LinSpecGCC} features, the GCC-PHAT channels are not shifted. The frequency shifting augmentation technique increases the variation of frequency patterns of sound events.  

\section{Experimental Settings}
\label{sec:exp}

\subsection{Dataset}
\label{sec:dataset}

The main dataset used in the majority of our experiments is the TNSSE 2021 dataset~\cite{Politis2021}. Since this dataset is relatively new, we also use the TNSSE 2020 dataset~\cite{Politis2020ADetection} to compare our models with state-of-the-art methods. The development subset of each TNSSE dataset consists of \num{400}, \num{100}, and \num{100} one-minute audio recordings for the train, validation, and test split, respectively. The evaluation subset of each dataset consists of \num{200} one-minute audio recording. Unless otherwise stated, the validation set was used for model selection while the test set was used for evaluation. Table~\ref{tab:datasets} summarizes some key characteristics of the two datasets. The azimuth and elevation ranges of both datasets are $[\SI{-180}{\degree}, \SI{180}{\degree})$ and $[\SI{-45}{\degree}, \SI{45}{\degree}]$, respectively.  

\begin{table}[t]
    \centering
    \caption {Characteristics of TNSSE 2020 and 2021 datasets}  
    \label{tab:datasets}
    \noindent\begin{tabularx}{\columnwidth}{lCC}
    \toprule
    Characteristics & 2020 & 2021 \\ 
    \midrule
    Channel format                  & FOA, MIC   & FOA, MIC \\
    Moving sources                  & \X    & \X  \\
    Ambiance noise                  & \X    & \X  \\
    Reverberation                   & \X    & \X  \\
    Unknown interferences           & \V    & \X  \\
    Maximum degree of polyphony     & 2     & 3   \\ 
    Number of target sound classes  & 14    & 12  \\
    \bottomrule
    \end{tabularx}
\end{table}

Both TNSSE datasets were recorded using a \num{32}-microphone Eigenmike spherical array with a radius of \SI{4.2}{\centi\meter}. The \num{32}-channel signals were converted into FOA format, whose array response is approximately frequency-independent up to around \SI{9}{\kilo\hertz}. Therefore, we compute EIV for SALSA features between \SI{50}{\hertz} and \SI{9}{\kilo\hertz}. Out of the \num{32} microphones, four microphones that form a tetrahedron are used for the MIC format. Since the radius of the spherical array corresponds to an aliasing frequency of \SI{4}{\kilo\hertz}, we computed EPV for MIC format between \SI{50}{\hertz} and \SI{4}{\kilo\hertz}. Even though the microphones are mounted on an acoustically-hard spherical baffle, we found that the far-field array model in Section~\ref{subsec:epv} is sufficient to extract the spatial cues for the MIC format. 

\subsection{Evaluation}

To evaluate SELD performance, we used the official evaluation metrics~\cite{Politis2020Overview2019} that were introduced in the 2021 DCASE Challenge as our default metrics. 
A sound event is considered a correct detection only if it has correct class prediction and its estimated DOA is less than $D\si{\degree}$ away from the DOA ground truth, where $D=\SI{20}{\degree}$ is the most commonly used value. The DOAE metrics are class-dependent, that is, the detected sound class will have to be correct in order for the corresponding localization predictions to count. Since some state-of-the-art SELD systems only reported the 2020 version of the DCASE evaluation metrics~\cite{Mesaros2019JointEvents}, we also used these metrics in some experiments to fairly compare the results.

Both the 2020 and 2021 SELD evaluation metrics consist of four metrics: location-dependent error rate ($\text{ER}_{\le \SI{20}{\degree}}$) and F1 score ($\text{F}_{\le \SI{20}{\degree}}$) for SED; and class-dependent localization error ($\text{LE}_\text{CD}$), and localization recall ($\text{LR}_\text{CD}$) for DOAE. We also computed an aggregated SELD error metric that was used as the ranking metric for the 2019 and 2020 DCASE Challenges as follows, 
\begin{align}
    \mathcal{E}_\textsc{seld} = \dfrac{1}{4}\left[\text{ER}_{\le \SI{20}{\degree}} + (1-\text{F}_{\le \SI{20}{\degree}}) + \dfrac{\text{LE}_\textsc{cd}}{\SI{180}{\degree}} + (1-\text{LR}_\textsc{cd})\right].
\end{align}
$\mathcal{E}_\text{SELD}$ was used for model and hyperparameter selection. A good SELD system should have low $\text{ER}_{\le \SI{20}{\degree}}$, high $\text{F}_{\le \SI{20}{\degree}}$, low $\text{LE}_\text{CD}$, high $\text{LR}_\text{CD}$, and low aggregated error metric $\mathcal{E}_\text{SELD}$.

\subsection{Hyperparameters}

We used a sampling rate of \SI{24}{\kilo\hertz}, window length of \SI{512}{ samples}, hop length of \SI{300}{samples}, Hann window, \num{512} FFT points, and \num{128} mel bands. As a result, the input frame rate of all the features was \SI{80}{fps}. Since the model temporally downsampled the input by a factor of \num{16}, we temporally upsampled the final outputs by a factor of \num{2} to match the label frame rate of \SI{10}{fps}. To reduce the feature dimensions to speed up the training time, we linearly compressed frequency bands above \SI{9}{\kilo\hertz}, which correspond to frequency bin index \num{192} and above, by a factor of \num{8}, \ie \num{8} consecutive bands will be averaged into a single band. As the results, the frequency dimension is $F=200$ for all linear-scale features. Unless stated otherwise, \num{8}-second audio chunks were used for model training. The loss weights for SED and DOAE were set to $\lambda=0.3$ and $\gamma=0.7$, respectively. Adam optimizer was used for all training. The learning rate was initially set to \num{3e-4} and linearly decreased to \num{e-4} over last \num{15} epochs of the total \num{50} training epochs. A threshold of \num{0.3} was used to binarize active class predictions in the SED outputs. 
\section{Results and discussion}
\label{sec:results}

We performed a series of experiments to compare the performances of each input feature with and without data augmentation. Afterwards, the effect of data augmentation on each feature was examined in details. We analyzed the effects of the magnitude and coherence tests on the performance of SELD systems running on SALSA features. Next, we studied the feature importance of \textsc{LinSpec}, EIV and EPV that constitute SALSA features. In addition, effect of different segment lengths on SALSA performance was investigated. For the MIC format, we examined effect of spatial aliasing on the SELD performance with SALSA features. Finally, we compared the performance of models trained on the proposed SALSA features with several state-of-the-art SELD systems on both the 2020 and 2021 TNSSE datasets. 

\subsection{Comparison between SALSA and other SELD features}

\begin{table*}[t]
    \centering
    \caption{Baseline SELD performances of different features without data augmentation.}
    \noindent\begin{tabularx}{\textwidth}{Xl rrrrr rrrrr}
    \toprule 
        \multirow{2}[2]{*}{Feature} & 
        \multirow{2}[2]{*}{Data Aug.} &  
        \multicolumn{5}{c}{FOA format} &
        \multicolumn{5}{c}{MIC format}
    \\ \cmidrule(lr){3-7}\cmidrule(lr){8-12}
        & & 
        $\downarrow$ $\text{ER}_{\le \SI{20}{\degree}}$ &
        $\uparrow$ $\text{F}_{\le \SI{20}{\degree}}$ &
        $\downarrow$ $\text{LE}_\text{CD}$&
        $\uparrow$ $\text{LR}_\text{CD}$ &
        $\downarrow$ $\mathcal{E}_\text{SELD}$ &
        $\downarrow$ $\text{ER}_{\le \SI{20}{\degree}}$ &
        $\uparrow$ $\text{F}_{\le \SI{20}{\degree}}$ &
        $\downarrow$ $\text{LE}_\text{CD}$&
        $\uparrow$ $\text{LR}_\text{CD}$ &
        $\downarrow$ $\mathcal{E}_\text{SELD}$
        \\ \midrule
        \textsc{MelSpecIV}   & None 
                    & 0.555 & 0.584 & 15.9\si{\degree} & 0.625 & 0.358 
                    & - & - & - & - & - \\
        \textsc{LinSpecIV}   & None
                    & \bf{0.527} & \bf{0.609} & 15.6\si{\degree} & \bf{0.642} & \bf{0.341}
                    & - & - & - & - & - \\
        \textsc{MelSpecGCC}  & None
                    & - & - & - & - & - 
                    & 0.660 & 0.455 & 21.1\si{\degree} & 0.521 & 0.450 \\ 
        \textsc{LinSpecGCC}  & None
                    & - & - & - & - & - 
                    & 0.622 & 0.506 & 19.6\si{\degree} & 0.583 & 0.410 \\
        \midrule
        FOA SALSA       & None
                    & 0.543 & 0.592 & \bf{15.4\si{\degree}} & 0.627 & 0.352 
                    & - & - & - & - & -  \\
        MIC SALSA       & None
                    & - & - & - & - & - 
                    & \bf{0.528} & \bf{0.601} & \bf{15.9\si{\degree}} & \bf{0.644} & \bf{0.343} \\
    \bottomrule
    \end{tabularx}
    \label{tab:baseline_noaug}
\end{table*}

\begin{table*}[t]
    \centering
    \caption{SELD performances of different features with best combination of data augmentation techniques.}
    \noindent\begin{tabularx}{\textwidth}{Xl rrrrr rrrrr }
    \toprule 
        \multirow{2}[2]{*}{Feature} & 
        \multirow{2}[2]{*}{Data Aug.} &  
        \multicolumn{5}{c}{FOA format} &
        \multicolumn{5}{c}{MIC format}
    \\ \cmidrule(lr){3-7}\cmidrule(lr){8-12}
        & & 
        $\downarrow$ $\text{ER}_{\le \SI{20}{\degree}}$ &
        $\uparrow$ $\text{F}_{\le \SI{20}{\degree}}$ &
        $\downarrow$ $\text{LE}_\text{CD}$&
        $\uparrow$ $\text{LR}_\text{CD}$ &
        $\downarrow$ $\mathcal{E}_\text{SELD}$ &
        $\downarrow$ $\text{ER}_{\le \SI{20}{\degree}}$ &
        $\uparrow$ $\text{F}_{\le \SI{20}{\degree}}$ &
        $\downarrow$ $\text{LE}_\text{CD}$&
        $\uparrow$ $\text{LR}_\text{CD}$ &
        $\downarrow$ $\mathcal{E}_\text{SELD}$
        \\ \midrule
        \textsc{MelSpecIV}   & CS + FS 
                    & 0.444 & 0.686 & 11.8\si{\degree} & 0.686 & 0.284 
                    & - & - & - & - & - \\
        \textsc{LinSpecIV}   & CS + FS + RC
                    & 0.410 & 0.710 & \bf{10.5\si{\degree}} & 0.702 & 0.264
                    & - & - & - & - & - \\
        \textsc{MelSpecGCC}  & CS + FS + RC
                    & - & - & - & - & - 
                    & 0.507 & 0.614 & 17.9\si{\degree} & 0.679 & 0.328 \\ 
        \textsc{LinSpecGCC}  & CS + FS + RC
                    & - & - & - & - & - 
                    & 0.514 & 0.606 & 17.8\si{\degree} & 0.676 & 0.333 \\
        \midrule                    
        FOA SALSA   & CS + FS
                    & \bf{0.404} & \bf{0.724} & 12.5\si{\degree} & \bf{0.727} & \bf{0.255} 
                    & - & - & - & - & - \\
        MIC SALSA   & CS + FS + RC
                    & - & - & - & - & - 
                    & \bf{0.408} & \bf{0.715} & \bf{12.6\si{\degree}} & \bf{0.728} & \bf{0.259} \\
    \bottomrule
    \multicolumn{8}{l}{CS: channel swapping; FS: frequency shifting; RC: random cutout.}
    \end{tabularx}
    \label{tab:bestaug}
\end{table*}

\Cref{tab:baseline_noaug} shows benchmark performances of all considered features without data augmentation. Linear-scale features (\textsc{LinSpec}-based) appear to perform better than their mel-scale counterparts (\textsc{MelSpec}-based) for both audio formats. 
For the `traditional' features, the performance gap between the FOA and MIC formats is large, with both IV-based features outperforming GCC-based features. Without data augmentation, FOA SALSA performed better than \textsc{MelSpecIV} but slightly worse than \textsc{LinSpecIV}, while MIC SALSA performed much better than both GCC-based features. 

\Cref{tab:bestaug} shows the performance of all features with their respective best combination of the three data augmentation techniques investigated. For the FOA format, the experimental results, again, showed that linear-scale features achieved better performance than mel-scale features. For the MIC format, the mel-scale features performed slightly better than linear-scale features. The large performance gap between the FOA and MIC formats still remained with data augmentation applied. IV-based features significantly outperform GCC-based features across all the evaluation metrics. With data augmentation, the proposed SALSA features achieved the best overall performances for both the FOA and MIC formats. SALSA scored the highest in \Fone and \LR; and the lowest in \ER and \Eseld among the setups investigated in \protect\Cref{tab:bestaug}. It is expected that a high \LR often leads to a high \LE. With a higher \LR, SALSA also has a higher \LE than \textsc{LinSpecIV} by \SI{2}{\degree}. SALSA outperformed both GCC-based features by a large margin. Compared to \textsc{MelSpecGCC}, SALSA feature substantially reduced \ER by \SI{20}{\percent}, increased \Fone by \SI{16}{\percent}, reduced \LE by \SI{5.3}{\degree}, and increased \LR by \SI{7}{\percent}. The overall \Eseld was impressively reduced by \SI{21}{\percent}. 

The performance gap between the IV- and GCC-based features, and the similar performances of SALSA for both array formats indicated that the exact TF mapping between the signal power and the directional cues, as per SALSA, \textsc{MelSpecIV}, and \textsc{LinSpecIV}, are much better for SELD than simply stacking spectrograms and GCC-PHAT spectra as per \textsc{MelSpecGCC} and \textsc{LinSpecGCC}. This exact TF mapping also facilitates the learning of CNNs, as the filters can more conveniently learn the multichannel local patterns on the image-like input features. Most importantly, the results showed that the extracted spatial cues for SALSA features are effective for both FOA and MIC formats. 
Therefore, SALSA can be considered as a unified SELD feature regardless of the array format. The outstanding performance gains in models trained with SALSA features shown in both \Cref{tab:baseline_noaug} and \ref{tab:bestaug} indicate that SALSA as a very effective feature for deep learning-based SELD. 

\subsection{Effect of data augmentation}

We report the effect of different data augmentation techniques on each feature in \Cref{tab:melspeciv_aug}. The experimental results clearly demonstrated that channel swapping significantly improved the performance for all features across all metrics. On average, \ER decreased by \SI{14.8}{\percent}, \Fone increased by \SI{13.8}{\percent}, \LE decreased by \SI{2.3}{\degree}, and \LR increased by \SI{7.8}{\percent}. Channel swapping reduced the aggregated error metric \Eseld by between \SI{13}{\percent} and \SI{16}{\percent}, where the larger reductions are observed for MIC features such as \textsc{MelSpecGCC}, \textsc{LinSpecGCC}, and MIC SALSA.  

When frequency shifting was used together with channel swapping, the performance was improved further for all features. Compared to channel swapping alone, the combination of channel swapping and frequency shifting on average reduced \ER by a further \SI{7.3}{\percent}, increased \Fone by \SI{5.8}{\percent}, reduced \LE by \SI{1.2}{\degree}, increased \LR by \SI{5.3}{\percent}, and reduced \Eseld by \SI{8.6}{\percent}. These results showed that varying the SED and DOA patterns by frequency shifting and channel swapping helped the models learn more effectively. 

When random cutout was used together with channel swapping and frequency shifting, the performance was further improved for \textsc{LinSpecIV} and all MIC features; but not for \textsc{MelSpecIV} and FOA SALSA. For subsequent experiments, the best combinations of data augmentation techniques for each feature, as shown in boldface in \Cref{tab:melspeciv_aug}, are used.

\begin{table}[t]
    \setlength\tabcolsep{3pt}
    \centering
    \caption{Performance of \textsc{MelSpecIV}, \textsc{LinSpecIV}, \textsc{MelSpecGCC}, \textsc{LinSpecGCC}, and SALSA with different data augmentation.}
    \noindent\begin{tabularx}{\columnwidth}{Xrrrrr}
    \toprule 
        Data Aug. 
        & $\downarrow \text{ER}_{\le \SI{20}{\degree}}$ 
        & $\uparrow \text{F}_{\le \SI{20}{\degree}}$
        & $\downarrow \text{LE}_\text{CD}$
        & $\uparrow \text{LR}_\text{CD}$ 
        & $\downarrow \mathcal{E}_\text{SELD}$ \\
    \midrule
    \bfseries \textsc{MelSpecIV}\\
        None        & 0.555 & 0.584 & 15.9\si{\degree} & 0.625 & 0.358 \\
        CS          & 0.472 & 0.655 & 12.0\si{\degree} & 0.653 & 0.308 \\
        \bf{CS+FS}  & 0.444 & \bf{0.686} & 11.8\si{\degree} & \bf{0.686} & \bf{0.284} \\
        CS+FS+RC    & \bf{0.440} & 0.683 & \bf{10.2\si{\degree}} & 0.668 & 0.286 \\
    \midrule
    \bfseries \textsc{LinSpecIV}\\
        None            & 0.527 & 0.609 & 15.6\si{\degree} & 0.642 & 0.341 \\
        CS              & 0.459 & 0.669 & 12.3\si{\degree} & 0.678 & 0.295 \\
        CS+FS           & 0.423 & 0.700 & 10.8\si{\degree} & 0.692 & 0.273 \\
        \bf{CS+FS+RC}   & \bf{0.410} & \bf{0.710} & \bf{10.5\si{\degree}} & \bf{0.702} & \bf{0.264} \\
    \midrule
    \bfseries FOA SALSA\\
    
        None            & 0.543 & 0.592 & 15.4\si{\degree} & 0.627 & 0.352 \\
        SC              & 0.462 & 0.655 & 14.9\si{\degree} & 0.666 & 0.306 \\
        \bf{CS+FS}      & \bf{0.404} & \bf{0.724} & 12.5\si{\degree} & \bf{0.727} & \bf{0.255} \\
        CS+FS+RC        & 0.413 & 0.713 & \bf{11.5\si{\degree}} & 0.713 & 0.263 \\
    \midrule\midrule
    \bfseries \textsc{MelSpecGCC}\\
    
        None            & 0.660 & 0.455 & 21.1\si{\degree} & 0.521 & 0.450 \\
        CS              & 0.552 & 0.556 & 18.1\si{\degree} & 0.583 & 0.378 \\
        CS+FS           & \bf{0.507} & 0.609 & \bf{17.0\si{\degree}} & 0.646 & 0.337 \\
        \bf{CS+FS+RC}   & \bf{0.507} & \bf{0.614} & 17.9\si{\degree} & \bf{0.679} & \bf{0.328} \\
    \midrule
    \bfseries \textsc{LinSpecGCC}\\
    
        None            & 0.622 & 0.506 & 19.6\si{\degree} & 0.583 & 0.410 \\
        CS              & 0.532 & 0.589 & 18.6\si{\degree} & 0.658 & 0.347 \\
        CS+FS           & \bf{0.514} & 0.604 & \bf{17.7\si{\degree}} & 0.666 & 0.336 \\
        \bf{CS+FS+RC}   & \bf{0.514} & \bf{0.606} & 17.8\si{\degree} & \bf{0.676} & \bf{0.333} \\
    \midrule
    \bfseries MIC SALSA\\
    
        None            & 0.528 & 0.601 & 15.9\si{\degree} & 0.644 & 0.343 \\
        CS              & 0.447 & 0.675 & 13.7\si{\degree} & 0.683 & 0.291 \\
        CS+FS           & 0.431 & 0.696 & \bf{12.3\si{\degree}} & 0.709 & 0.274 \\
        \bf{CS+FS+RC}   & \bf{0.408} & \bf{0.715} & 12.6\si{\degree} & \bf{0.728} & \bf{0.259} \\
    \bottomrule
    \multicolumn{6}{l}{CS: channel swapping; FS: frequency shifting; RC: random cutout.}
    \end{tabularx}
    \label{tab:melspeciv_aug}
    \label{tab:linspeciv_aug}
    \label{tab:melspecgcc_aug}
    \label{tab:linspecgcc_aug}
    \label{tab:salsa_foa_aug}
    \label{tab:salsa_mic_aug}
\end{table}

\subsection{Effect of magnitude and coherence tests}

\begin{table}[t]
    \setlength\tabcolsep{3pt}
    \centering
    \caption{Effect of magnitude and coherence tests on SALSA features.}
    \noindent\begin{tabularx}{\columnwidth}{Xrrrrr}
    \toprule 
        Test&
        $\downarrow$ $\text{ER}_{\le \SI{20}{\degree}}$ &
        $\uparrow$ $\text{F}_{\le \SI{20}{\degree}}$ &
        $\downarrow$ $\text{LE}_\text{CD}$&
        $\uparrow$ $\text{LR}_\text{CD}$ &
        $\downarrow$ $\mathcal{E}_\text{SELD}$
        \\ \midrule
       \bfseries FOA SALSA\\
        
        None    & 0.418 & 0.706 & 12.0\si{\degree} & 0.710 & 0.267\\
        Magnitude    & 0.434 & 0.698 & \bf{11.9\si{\degree}} & 0.701 & 0.275\\
        Magnitude + Coherence
                & \bf{0.404} & \bf{0.724} & 12.5\si{\degree} & \bf{0.727} & \bf{0.255}\\
        \midrule
        \bfseries MIC SALSA\\
        
        None  & 0.414 & 0.701 & 12.1\si{\degree} & 0.700 & 0.270 \\
        Magnitude    & \bf{0.407} & \bf{0.716} & \bf{12.3\si{\degree}} & 0.721 & 0.260 \\
        Magnitude + Coherence  & 0.408 & 0.715 & 12.6\si{\degree} & \bf{0.728} & \bf{0.259} \\ 
    \bottomrule
    \multicolumn{6}{l}{CS: channel swapping; FS: frequency shifting; RC: random cutout.}
    \label{tab:sp_test}
    \end{tabularx}
\end{table}

\Cref{tab:sp_test} shows the effect of magnitude and coherence tests on the performance of models trained on SALSA features. \Cref{fig:tf_bin_dist} indicates that around \SI{33}{\percent} of all TF bins are removed after the magnitude test, and an additional \SI{20}{\percent} of bins are removed after the coherence test. These tests aim to only include approximately single-source TF bins with reliable directional cues. The magnitude test improved the performance of the MIC format but not the FOA format. On the other hand, using both the magnitude and coherence tests significantly improved the performance of the FOA format. Overall, when both tests are applied to compute SALSA features, the performances were improved compared to when no test was applied. For subsequent experiments, both tests were applied to compute SALSA features. 

\subsection{Feature importance}

\begin{table}[t]
    \centering
    \setlength\tabcolsep{3pt}
    \caption{Feature importance of FOA and MIC SALSA.}
    \noindent\begin{tabularx}{\columnwidth}{Xrrrrr}
    \toprule 
        Components
        & $\downarrow \text{ER}_{\le \SI{20}{\degree}}$ 
        & $\uparrow \text{F}_{\le \SI{20}{\degree}}$
        & $\downarrow \text{LE}_\text{CD}$
        & $\uparrow \text{LR}_\text{CD}$ 
        & $\downarrow \mathcal{E}_\text{SELD}$ \\
    \midrule
    \bfseries FOA SALSA\\
    
        \textsc{LinSpec}            & 0.835 & 0.123 & 87.2\si{\degree} & 0.608 & 0.647 \\
        EIV           & 0.577 & 0.557 & 14.1\si{\degree} & 0.571 & 0.382 \\
        \textsc{Mono-SALSA}             & 0.421 & 0.705 & 12.8\si{\degree} & 0.723 & 0.266 \\
        SALSA              & \bf{0.404} & \bf{0.724} & \bf{12.5\si{\degree}} & \bf{0.727} & \bf{0.255} \\
    \midrule
    \bfseries MIC SALSA\\
    
        \textsc{LinSpec}            & 0.506 & 0.616 & 18.1\si{\degree} & 0.698 & 0.323 \\
        EPV           & 0.629 & 0.502 & 17.4\si{\degree} & 0.547 & 0.419 \\
        \textsc{Mono-SALSA}             & 0.443 & 0.680 & 14.7\si{\degree} & 0.710 & 0.284 \\
        SALSA              & \bf{0.408} & \bf{0.715} & \bf{12.6\si{\degree}} & \bf{0.728} & \bf{0.259} \\
    \bottomrule
    \end{tabularx}
    \label{tab:salsa_foa_feature_imp}
    \label{tab:salsa_mic_feature_imp}
\end{table}

\Cref{tab:salsa_foa_feature_imp} reports the feature importance of each component in SALSA feature: multichannel log-linear spectrogram \textsc{LinSpec}, as well as spatial features EIV and EPV for FOA and MIC formats, respectively. \textsc{Mono-SALSA} is an ablation feature formed by stacking the log-linear spectrogram of only the first microphone with the corresponding spatial features.
For both formats, SALSA achieved the best performance, followed by \textsc{Mono-SALSA}. 

For the FOA format, \textsc{LinSpec} alone could not meaningfully estimate DOAs. One possible reason is that the spatial cues of FOA format are encoded in the signed amplitude differences between microphones, but \textsc{LinSpec} retains only the unsigned magnitude differences. 
The sign ambiguity caused the confusion between the input features and the target labels. Therefore, the model trained on \textsc{LinSpec} feature failed to detect the correct DOAs. On the other hand, the model trained on only the EIV feature performed reasonably well. The EIV feature preserved some coarse spatiotemporal patterns of each sound class (see \Cref{fig:foa_salsa}), thus the model was able to distinguish different sound classes. 
SALSA feature significantly outperformed its constituent features, \textsc{LinSpec} and EIV. In the absence of the X, Y, and Z channels of the linear spectrograms, \textsc{Mono-SALSA} performed slightly worse than SALSA on the SED metrics but similarly on the DOAE metrics. These results suggest that the main contribution of the X, Y, and Z channels in the linear spectrograms is to distinguish different sound classes. 

For the MIC format, \textsc{LinSpec} alone performed reasonably well for SELD. Referring to \Cref{sec:dataset}, the MIC format of the DCASE SELD dataset is not a true far-field array, but rather a baffled microphone array, where some spatial cues are also encoded in the magnitude differences between microphones. Therefore, not only is the model trained on \textsc{LinSpec} feature able to classify sound sources, but it is also able to estimate DOAs. The EPV feature alone returned a lower SELD performance compared to the EIV feature of the FOA format, likely because the EPV feature is computed with an upper cutoff frequency of \SI{4}{\kilo\hertz}, which is much lower than that of EIV at \SI{9}{\kilo\hertz}. 
The model trained on only EPV also has the highest \ER and lowest \Fone among all ablation models of the MIC format. SALSA feature significantly outperformed its individual feature component across all the metrics. The performance gap between SALSA and \textsc{Mono-SALSA} is larger for the MIC format than the FOA format. The reason is likely that the spatial cues are also encoded in the magnitude of different input channels, and the EPV is all zeroed out above the upper cutoff frequency. Therefore, the multichannel nature of the spectrograms play an important role in both sound class recognition and DOA estimation. 

\subsection{Effect of spatial aliasing on SELD for microphone array}

\begin{table}[t]
    \centering
    \setlength\tabcolsep{3pt}
    \caption{Effect of spatial aliasing on SALSA feature of MIC format.}
    \noindent\begin{tabularx}{\columnwidth}{Xrrrrr}
    \toprule 
        Cutoff frequency 
        & $\downarrow \text{ER}_{\le \SI{20}{\degree}}$ 
        & $\uparrow \text{F}_{\le \SI{20}{\degree}}$
        & $\downarrow \text{LE}_\text{CD}$
        & $\uparrow \text{LR}_\text{CD}$ 
        & $\downarrow \mathcal{E}_\text{SELD}$ \\
    \midrule
        \SI{2.0}{\kilo\hertz}          & \bf{0.403} & 0.714 & \bf{12.5\si{\degree}} & 0.707 & 0.261 \\
        \SI{4.0}{\kilo\hertz}          & 0.408 & \bf{0.715} & 12.6\si{\degree} & \bf{0.728} & \bf{0.259} \\
        \SI{9.0}{\kilo\hertz}          & 0.425 & 0.698 & 12.8\si{\degree} & 0.720 & 0.270 \\
    \bottomrule
    \end{tabularx}
    \label{tab:salsa_mic_aliasing}
\end{table}

For narrow band signals, spatial aliasing occurs at high frequency bins, where half of the signal wavelength is less than the distance between two microphones. 
To investigate the effect of spatial aliasing when SALSA features for MIC format are used, we report the performances of SALSA with different upper cutoff frequencies in \Cref{tab:salsa_mic_aliasing}. The upper cutoff frequencies were computed using the spatial aliasing formula for narrow band signals, $f_\text{alias}=c/(2d_\text{max})$, where $d_\text{max}$ is the maximum distance between any two microphones in the array. The investigated values of $d_\text{max}$ are the arc length between any two microphones (\SI{8.0}{\centi\meter}) and the radius of the Eigenmike array (\SI{4.2}{\centi\meter}), which correspond to aliasing frequencies of \SI{2}{\kilo\hertz}, and \SI{4}{\kilo\hertz}, respectively. In addition, we also tested a cutoff frequency of \SI{9}{\kilo\hertz} to investigate the case where spatial aliasing is ignored. \Cref{tab:salsa_mic_aliasing} shows that cutoff frequencies at \SI{2}{\kilo\hertz} and \SI{4}{\kilo\hertz} result in similar performances. One possible reason is that the spatial aliasing might not significantly occur in all of the microphone pairs beyond \SI{2}{\kilo\hertz} for some DOAs. On the other hand, with the \SI{9}{\kilo\hertz} cutoff frequency, spatial aliasing has occurred in too many high-frequency bins, resulting in a slightly lower performance than a loose cutoff frequency at \SI{4}{\kilo\hertz}. However, the impact of spatial aliasing appears to be mild, with the model trained on a loose aliasing frequency at \SI{4}{\kilo\hertz} achieving the best \Eseld. This result is agreeable with the finding in \cite{Dmochowski2009OnArrays}, where broadband signals were shown to not experience spatial aliasing unless they contain strong harmonic components. 

\subsection{Effect of segment length for training}

\begin{table}[t]
    \setlength{\tabcolsep}{3pt}
    \centering
    \caption{Effect of segment length during training on SELD performance using SALSA.}
    \noindent\begin{tabularx}{\columnwidth}{Xrrrrr}
    \toprule 
        Length &
        $\downarrow$ $\text{ER}_{\le \SI{20}{\degree}}$ &
        $\uparrow$ $\text{F}_{\le \SI{20}{\degree}}$ &
        $\downarrow$ $\text{LE}_\text{CD}$&
        $\uparrow$ $\text{LR}_\text{CD}$ &
        $\downarrow$ $\mathcal{E}_\text{SELD}$
        \\ \midrule
        \bfseries FOA SALSA\\
        
        \SI{4}{\second}      & 0.468 & 0.658 & 13.4\si{\degree} & 0.646 & 0.310 
                \\
        \SI{8}{\second}       & \bf{0.404} & \bf{0.724} & 12.5\si{\degree} & \bf{0.727} & \bf{0.255}
               \\
        \SI{12}{\second}      & 0.414 & 0.717 & \bf{11.8\si{\degree}} & 0.720 & 0.261
               \\ 
        \midrule
        \bfseries MIC SALSA\\
        
        \SI{4}{\second}      
                & 0.449 & 0.664 & 14.1\si{\degree} & 0.678 & 0.297 \\
        \SI{8}{\second}     
                & \bf{0.408} & \bf{0.715} & \bf{12.6\si{\degree}} & 0.728 & \bf{0.259} \\
        \SI{12}{\second}    & 0.413 & 0.714 & 12.7\si{\degree} & \bf{0.730} & 0.260 \\ 
    \bottomrule
    \end{tabularx}
    \label{tab:segment_length}
\end{table}

Different sound events often have different duration. Thus the segment length that is used during training may affect the model performance. The sound event lengths from the TNSSE 2021 dataset are between \SI{0.2}{\second} and \SI{40.0}{\second}, with a median of \SI{3.2}{\second}, and a mean of \SI{8.3}{\second}. We present the SELD performances on models trained with different input segment lengths, as per \Cref{tab:segment_length}. Models trained with a segment length of \SI{8}{\second} significantly outperformed models trained with a segment length of \SI{4}{\second} for both the FOA and MIC formats. However, increasing the segment length to \SI{12}{\second} did not further improve the overall performance. Thus, it appears that the model requires a certain minimum sequence length to sufficiently learn the temporal dependency, although this temporal dependency does not need to be very long, since the model would likely rely more on recent frames than older frames.  

\subsection{Comparisons with state-of-the-art methods for SELD}

We compared models trained with the proposed SALSA features with state-of-the-art (SOTA) methods on three datasets: the test and evaluation splits of the TNSSE 2020 dataset~\cite{Politis2020ADetection} and the test split of the TNSSE 2021 dataset~\cite{Politis2021}. Some of the SOTA methods used single models while others used ensemble models. We used the same single-model SELD network shown in \Cref{sec:net} to train all of the models reported, \ie{} no ensembling was used. To further improve the performance of our models, we applied test-time augmentation (TTA) during inference~\cite{Shimada2021ACCDOA:Detection}. TTA swaps the channels of the SALSA features in a manner similar to the channel swapping augmentation technique that was employed during training. The estimated DOA outputs were rotated back to the original axes, then averaged to produce the final results. During inference, the whole \num{60}-second features were passed into the models without being split into smaller chunks. Since the SOTA results on the TNSSE 2020 dataset were evaluated using the 2020 SELD evaluation metrics, we evaluated our models using both the 2020 and 2021 metrics, the former for fair comparison with past works, and the latter for ease of comparison with future works.  

\subsubsection{Performance on the test split of the TNSSE 2020 dataset}

\begin{table}[t] 
    \setlength{\tabcolsep}{3pt}
    \centering
    \caption{SELD performances of SOTA systems and SALSA-based models on test split of the TNSSE 2020 dataset.}
    \noindent\begin{tabularx}{\columnwidth}{Xlrrrr}
    \toprule 
        System & Format &
        $\text{ER}_{\le \SI{20}{\degree}}$ &
        $\text{F}_{\le \SI{20}{\degree}}$ &
        $\text{LE}_\text{CD}$&
        $\text{LR}_\text{CD}$ \\
    \midrule
    \bfseries 2020 Metrics \\
        DCASE baseline~\cite{Politis2020ADetection}
            & FOA & 0.72\hphantom{0} & 0.374 & 22.8\si{\degree} & 0.607 \\
        Shimada\etalcite{Shimada2021ACCDOA:Detection} w/o TTA
            & FOA & 0.36\hphantom{0} & 0.730 & 10.2\si{\degree} & 0.791 \\
        Shimada\etalcite{Shimada2021ACCDOA:Detection} w/ TTA
            & FOA &0.32\hphantom{0} & 0.768 & 7.9\si{\degree} & 0.805 \\
        Wang\etalcite{Wang2020TheChallenge}
            & FOA+MIC & 0.29\hphantom{0} & 0.764 & 9.4\si{\degree} & 0.828 \\
        ('20 \#1) Wang\etalcite{Wang2020TheChallenge} $\star$ 
            & FOA+MIC & \bf{0.26\hphantom{0}} & \bf{0.800} & \bf{7.4\si{\degree}} & \bf{0.847} \\
    \midrule
        FOA SALSA w/o TTA
            & FOA & 0.338 & 0.748 & 7.9\si{\degree} & 0.784 \\
        MIC SALSA w/o TTA
            & MIC & 0.379 & 0.717 & 8.2\si{\degree} & 0.762 \\ 
        FOA SALSA w/ TTA
            & FOA & 0.318 & 0.761 & 7.4\si{\degree} & 0.797 \\
        MIC SALSA w/ TTA
            & MIC & 0.341 & 0.741 & 7.8\si{\degree} & 0.783 \\
    \midrule
    \midrule
    \bfseries 2021 Metrics \\
        FOA SALSA w/o TTA
            & FOA & 0.344 & 0.755 & 8.1\si{\degree} & 0.755 \\
        MIC SALSA w/o TTA
            & MIC & 0.383 & 0.727 & 8.3\si{\degree} & 0.738 \\ 
        FOA SALSA w/ TTA
            & FOA & 0.323 & 0.768 & 7.5\si{\degree} & 0.763 \\
        MIC SALSA w/ TTA
            & MIC & 0.342 & 0.749 & 7.9\si{\degree} & 0.744 \\
    \bottomrule
    \multicolumn{2}{l}{$\star$ denotes an ensemble model.}
    \end{tabularx}
    \label{tab:sota_2020_dev}
\end{table}

\Cref{tab:sota_2020_dev} shows the performances on the test split of the TNSSE 2020 dataset of SOTA systems, and our SALSA models for both the FOA and MIC formats. FOA SALSA models performed slightly better than the MIC counterparts. The TTA significantly improved location dependent SED metrics \ER and \Fone. The model by Wang\etalcite{Wang2020TheChallenge} used both the FOA and MIC data as input features and achieved the best performance for \ER, \Fone, and \LR for single models. However, it is considerably more expensive to have both FOA and MIC data available in real-life applications due to the more specialized recording setup required. Our FOA SALSA model outperformed the DCASE baseline~\cite{Politis2020ADetection} by a large margin, and performed better than \cite{Shimada2021ACCDOA:Detection} in term of \ER, \Fone, and \LE. Our FOA SALSA model with TTA also performed on-par with the TTA version of \cite{Shimada2021ACCDOA:Detection}. On average, the 2021 evaluation metrics return similar \ER, \Fone and \LE compared to the 2020 metrics, but stricter \LR than the 2020 metrics. 

\subsubsection{Performance on the evaluation split of the TNSSE 2020 dataset}

\begin{table}[t] \small

    \centering
    \caption{SELD performances of SOTA systems and SALSA-based models on evaluation split of TNSSE 2020 dataset.}
    \setlength{\tabcolsep}{3pt}
    \footnotesize
    \noindent\begin{tabularx}{\columnwidth}{Xlrrrr}
    \toprule 
        System
        & Format
        & $\text{ER}_{\le \SI{20}{\degree}}$ 
        & $\text{F}_{\le \SI{20}{\degree}}$
        & $\text{LE}_\text{CD}$
        & $\text{LR}_\text{CD}$ \\
    \midrule
    \bfseries 2020 Metrics \\
        DCASE'21 baseline~\cite{Politis2020ADetection}
            & MIC & 0.69\hphantom{0} & 0.413 & 23.1\si{\degree} & 0.624 \\
        Cao\etalcite{Cao2021AnDetection}
            & FOA & 0.233 & 0.832 & 6.8\si{\degree} & 0.861 \\
        ('20 \#2) Nguyen\etalcite{Nguyen2020EnsembleTracking} $\star$  
            & FOA & 0.23\hphantom{0} & 0.820 & 9.3\si{\degree} & \bf{0.900} \\
        ('20 \#1) Wang\etalcite{Wang2020TheChallenge} $\star$ 
            & FOA+MIC & \bf{0.20\hphantom{0}} & 0.849 & \bf{6.0\si{\degree}} & 0.885 \\
    \midrule
        FOA SALSA w/o TTA
            & FOA & 0.237 & 0.823 & 6.9\si{\degree} & 0.858 \\
        MIC SALSA w/o TTA
            & MIC & 0.227 & 0.836 & 6.7\si{\degree} & 0.869 \\ 
        FOA SALSA w/ TTA
            & FOA & 0.219 & 0.840 & 6.5\si{\degree} & 0.869 \\
        MIC SALSA w/ TTA
            & MIC & \bf{0.202} & \bf{0.854} & \bf{6.0\si{\degree}} & 0.884 \\
    \midrule
    \midrule
    \bfseries 2021 Metrics \\
        FOA SALSA w/o TTA
            & FOA & 0.244 & 0.830 & 7.0\si{\degree} & 0.831 \\
        MIC SALSA w/o TTA
            & MIC & 0.234 & 0.842 & 6.7\si{\degree} & 0.849 \\ 
        FOA SALSA w/ TTA
            & FOA & 0.225 & 0.844 & 6.6\si{\degree} & 0.838 \\
        MIC SALSA w/ TTA
            & MIC & 0.208 & 0.858 & 6.0\si{\degree} & 0.856 \\
    \bottomrule
    \multicolumn{2}{l}{$\star$ denotes an ensemble model.}
    \end{tabularx}
    \label{tab:sota_2020_eval}
\end{table}

\Cref{tab:sota_2020_eval} shows the performances on the evaluation split of the TNSSE 2020 dataset of SOTA systems, and our SALSA models for both the FOA and MIC formats. Our models were trained using all \num{600} audio clips from the development split of the TNSSE 2020 dataset. Interestingly, when more data are available for training, models trained on MIC SALSA features performed better than models trained on FOA SALSA features across all metrics. The FOA SALSA model has competitive performance compared to \cite{Cao2021AnDetection} while the MIC SALSA model performed slightly better. The MIC SALSA model with TTA achieved comparable performance as the top ensemble model~\cite{Wang2020TheChallenge} from the 2020 DCASE Challenge, with similar \ER, \LE, \LR and higher \Fone. The 2021 metrics again returned similar \ER, \Fone, \LE results and stricter \LR than the 2020 metrics.   

\subsubsection{Performance on the test split of the TNSSE 2021 dataset}

\begin{table}[t] \small
    \centering
    \caption{SELD performances of SOTA systems and SALSA-based models on test split of TNSSE 2021 dataset.}
    
    \setlength{\tabcolsep}{3pt}
    \footnotesize
    \noindent\begin{tabularx}{\columnwidth}{Xlrrrrr}
    \toprule 
        System
        & Format
        & $\text{ER}_{\le \SI{20}{\degree}}$ 
        & $\text{F}_{\le \SI{20}{\degree}}$
        & $\text{LE}_\text{CD}$
        & $\text{LR}_\text{CD}$ \\
    \midrule
    \bfseries 2021 Metrics \\
        DCASE baseline~\cite{Politis2021}
            & FOA & 0.73\hphantom{0} & 0.307 & 24.5\si{\degree} & 0.448 \\
        ('21 \#1) Shimada\etalcite{Shimada2021EnsembleDetection} $\star$
            & FOA & 0.43\hphantom{0} & 0.699 & \bf{11.1\si{\degree}} & 0.732 \\
        ('21 \#4) Lee\etalcite{Lee2021SoundChallenge} $\star$
            & FOA & 0.46\hphantom{0} & 0.609 & 14.4\si{\degree} & 0.733 \\
    \midrule
        FOA SALSA w/o TTA
            & FOA & 0.404 & 0.724 & 12.5\si{\degree} & 0.727 \\
        MIC SALSA w/o TTA
            & MIC & 0.408 & 0.715 & 12.6\si{\degree} & 0.728 \\ 
        FOA SALSA w/ TTA
            & FOA & 0.376 & \bf{0.744} & \bf{11.1\si{\degree}} & 0.722 \\
        MIC SALSA w/ TTA
            & MIC & 0.376 & 0.735 & 11.2\si{\degree} & 0.722 \\ 
        ('21 \#2) Nguyen\etalcite{Nguyen2021DCASEDetection} $\star$
            & FOA & \bf{0.37\hphantom{0}} & 0.737 & 11.2\si{\degree} & \bf{0.741} \\
    \bottomrule
    \multicolumn{2}{l}{$\star$ denotes an ensemble model.}
    \end{tabularx}
    \label{tab:sota_2021_dev}
\end{table}

\Cref{tab:sota_2021_dev} shows the performances on the test split of the TNSSE 2021 dataset of SOTA systems, and our SALSA models for both the FOA and MIC formats. The FOA SALSA models performed similarly in \ER, \LE, \LR as and higher \Fone compared to the MIC SALSA models. The TTA significantly improved their \ER, \Fone, and \LE but not \LR. The models trained on SALSA features of both formats outperformed the DCASE baseline by the large margin, and performed better than the highest-ranked system from the 2021 DCASE Challenge \cite{Shimada2021EnsembleDetection} in terms of \ER and \Fone. With TTA, the models trained on SALSA features achieved much better \ER and \Fone, similar \LE, and slightly lower \LR compared to \cite{Shimada2021EnsembleDetection}. An ensemble model trained on a variant of our proposed SALSA features~\cite{Nguyen2021DCASEDetection} officially ranked second in the team category of the SELD taks in the 2021 DCASE Challenge. The SALSA variant in \cite{Nguyen2021DCASEDetection} included an additional channel for the estimated DRR at each TF bin. 

Compared to the TNSSE 2020 dataset, the TNSSE 2021 dataset is more challenging since it has more overlapping sound events and unknown directional interferences. Overall, the performances of models listed in \Cref{tab:sota_2021_dev} are lower than those of the models listed in \Cref{tab:sota_2020_dev} across all metrics.

The results in Tables \ref{tab:sota_2020_dev} to \ref{tab:sota_2021_dev} consistently show that the proposed SALSA features for both the FOA and MIC formats are very effective for SELD. Simple CRNN models trained on SALSA features surpassed or performed comparably to many SOTA systems, both single models and ensembles, on different datasets across all evaluation metrics.

\subsection{Qualitative evaluation}

\begin{figure}[tb]
    \centering
    \includegraphics[width=\columnwidth]{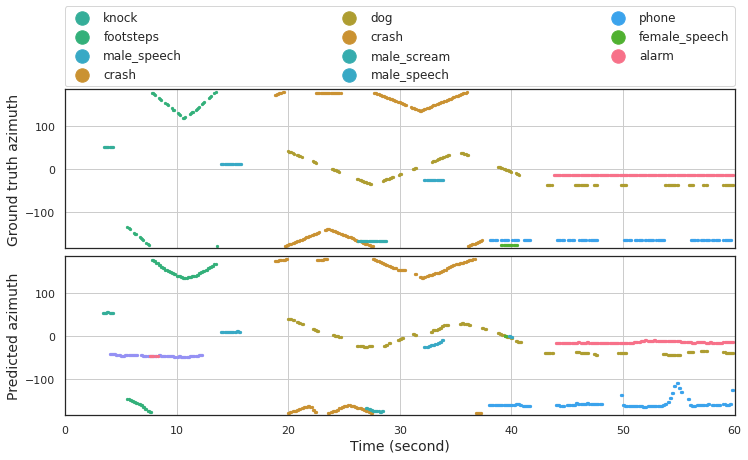}
    \vspace{-0.5cm}
    \caption{Visualization of ground truth and predicted azimuth for test clip \texttt{fold6\_room2\_mix041} of the TNSSE 2021 dataset. Legend lists the ground truth events in chronological order. Sound classes are color-coded. 
    {\textsc{piano} event (purple) and an \textsc{alarm} event (pink) were misclassified between the \num{4}\textsuperscript{th} and the \num{12}\textsuperscript{th} seconds}}
    \label{fig:gt_pred_visualization}
\end{figure} 

\Cref{fig:gt_pred_visualization} shows the plots of ground truth and predicted azimuth angles for a audio clip from the test set of the TNSSE 2021 dataset. The angles were predicted by a CRNN model trained on FOA SALSA features. Overall, the trajectories of predicted events were smooth and followed the ground truths closely. The model was able to correctly detect the sound classes and estimate DOAs across different numbers of overlapping sound sources (up to three overlapping sources). An unknown interference was misclassified as a \textsc{piano} event (purple) and an \textsc{alarm} event (pink) between the \num{4}th and the \num{12}th seconds. Since we used the class-wise output format to train the model, when there were two overlapping \textsc{crash} events between the \num{22}nd and the \num{24}th seconds, the model only predicted one \textsc{crash} event.
\section{Conclusion}
\label{sec:conclusion}

In conclusion, we proposed a novel and effective feature for polyphonic SELD named \emph{Spatial cue-Augmented Log-SpectrogrAm} (SALSA), which consists of multichannel log-spectrograms and normalized principal eigenvector of the spatial covariance matrix at each TF bin of the spectrograms. There are two key characteristics that contribute to the effectiveness of the proposed feature. Firstly, SALSA spectrotemporally aligns the signal power and the source directional cues, which aids in resolving overlapping sound sources. This locally linear alignment works well with CNNs, where the filters learn the multichannel local pattern of the image-like input features. Secondly, SALSA includes helpful directional cues extracted from the principal eigenvectors of the spatial covariance matrices. Depending on the array type, where the directional cues might be encoded as interchannel amplitude and/or phase differences, the principal eigenvectors can be easily normalized to extract these cues. Therefore, SALSA features are versatile to use with different microphone array formats, such as FOA and MIC. 

The proposed SALSA features can be further enhanced by incorporating signal processing-based methods such as magnitude and coherence tests to select more reliable directional cues and improve SELD performance. In addition, for multichannel arrays, spatial aliasing has little effect on the performance of models trained on SALSA. More importantly, the training segment length must be sufficient long for the model to capture the temporal dependency in the data.   

In addition, data augmentation techniques such as channel swapping, frequency shifting, and random cutout can be readily applied to SALSA on the fly during training. These data augmentation techniques mitigated the problem of small datasets and significantly improved the performance of models trained on SALSA features. Simple CRNN models trained on the SALSA features achieved similar or even better SELD performance than many state-of-the-art systems on the TNSSE 2020 and 2021 datasets. 

\bibliographystyle{packages/IEEEtran}
\bibliography{references}

\begin{thebibliography}{10}
\providecommand{\url}[1]{#1}
\def\UrlFont{\rmfamily}
\providecommand{\newblock}{\relax}
\providecommand{\bibinfo}[2]{#2}
\providecommand\BIBentrySTDinterwordspacing{\spaceskip=0pt\relax}
\providecommand\BIBentryALTinterwordstretchfactor{4}
\providecommand\BIBentryALTinterwordspacing{\spaceskip=\fontdimen2\font plus
\BIBentryALTinterwordstretchfactor\fontdimen3\font minus
  \fontdimen4\font\relax}
\providecommand\BIBforeignlanguage[2]{{%
\expandafter\ifx\csname l@#1\endcsname\relax
\typeout{** WARNING: IEEEtran.bst: No hyphenation pattern has been}%
\typeout{** loaded for the language `#1'. Using the pattern for}%
\typeout{** the default language instead.}%
\else
\language=\csname l@#1\endcsname
\fi
#2}}

\bibitem{Salamon2017DeepClassification}
J.~Salamon and J.~P. Bello, ``{Deep Convolutional Neural Networks and Data
  Augmentation for Environmental Sound Classification},''
  \emph{\protect\JournalTitle{IEEE Signal Processing Letters}}, vol.~24, no.~3,
  pp. 279--283, 2017.

\bibitem{Stowell2016BirdChallenge}
D.~Stowell, M.~Wood, Y.~Stylianou, and H.~Glotin, ``{Bird detection in audio: A
  survey and a challenge},'' in \emph{\protect\JournalTitle{Proceedings of the
  IEEE International Workshop on Machine Learning for Signal Processing}},
  2016.

\bibitem{Foggia2016AudioSounds}
P.~Foggia, N.~Petkov, A.~Saggese, N.~Strisciuglio, and M.~Vento, ``{Audio
  Surveillance of Roads: A System for Detecting Anomalous Sounds},''
  \emph{\protect\JournalTitle{IEEE Transactions on Intelligent Transportation
  Systems}}, vol.~17, no.~1, pp. 279--288, 2016.

\bibitem{Valin2004LocalizationApproach}
J.~M. Valin, F.~Michaud, B.~Hadjou, and J.~Rouat, ``{Localization of
  simultaneous moving sound sources for mobile robot using a frequency-domain
  steered beamformer approach},'' in \emph{\protect\JournalTitle{Proceedings of
  the IEEE International Conference on Robotics and Automation}}, 2004, pp.
  1033--1038.

\bibitem{Adavanne2019SoundNetworks}
S.~Adavanne, A.~Politis, J.~Nikunen, and T.~Virtanen, ``{Sound Event
  Localization and Detection of Overlapping Sources Using Convolutional
  Recurrent Neural Networks},'' \emph{\protect\JournalTitle{IEEE Journal of
  Selected Topics in Signal Processing}}, vol.~13, no.~1, pp. 34--48, 2019.

\bibitem{Hirvonen2015ClassificationNetworks}
T.~Hirvonen, ``{Classification of spatial audio location and content using
  Convolutional neural networks},'' in \emph{\protect\JournalTitle{Proceedings
  of the 138th Audio Engineering Society Convention 2015}}, 2015, pp. 622--631.

\bibitem{Cao2019PolyphonicStrategy}
Y.~Cao, Q.~Kong, T.~Iqbal, F.~An, W.~Wang, and M.~D. Plumbley, ``{Polyphonic
  Sound Event Detection and Localization using a Two-Stage Strategy},'' in
  \emph{\protect\JournalTitle{Proceedings of the 4th Workshop on Detection and
  Classification of Acoustic Scenes and Events}}, 2019.

\bibitem{Mazzon2019FirstEstimation}
L.~Mazzon, Y.~Koizumi, M.~Yasuda, and N.~Harada, ``{First Order Ambisonics
  Domain Spatial Augmentation for DNN-based Direction of Arrival Estimation},''
  in \emph{\protect\JournalTitle{Proceedings of the 4th Workshop on Detection
  and Classification of Acoustic Scenes and Events}}, 2019, pp. 154--158.

\bibitem{Nguyen2020ADetection}
T.~N.~T. Nguyen, D.~L. Jones, and W.~Gan, ``{A Sequence Matching Network for
  Polyphonic Sound Event Localization and Detection},'' in
  \emph{\protect\JournalTitle{Proceedings of the IEEE International Conference
  on Acoustics, Speech and Signal Processing}}, 2020, pp. 71--75.

\bibitem{Nguyen2021ANetwork}
T.~N.~T. Nguyen, N.~K. Nguyen, H.~Phan, L.~Pham, K.~Ooi, D.~L. Jones, and W.-S.
  Gan, ``{A General Network Architecture for Sound Event Localization and
  Detection Using Transfer Learning and Recurrent Neural Network},'' in
  \emph{\protect\JournalTitle{Proceedings of the IEEE International Conference
  on Acoustics, Speech and Signal Processing}}, 2021, pp. 935--939.

\bibitem{Cao2020Event-independentDetection}
Y.~Cao, T.~Iqbal, Q.~Kong, Y.~Zhong, W.~Wang, and M.~D. Plumbley,
  ``{Event-independent Network for Polyphonic Sound Event Localization and
  Detection},'' in \emph{\protect\JournalTitle{Proceedings of the 5th Workshop
  on Detection and Classification of Acoustic Scenes and Events}}, 2020.

\bibitem{Cao2021AnDetection}
Y.~Cao, T.~Iqbal, Q.~Kong, F.~An, W.~Wang, and M.~D. Plumbley, ``{An Improved
  Event-Independent Network for Polyphonic Sound Event Localization and
  Detection},'' in \emph{\protect\JournalTitle{Proceedings of the IEEE
  International Conference on Acoustics, Speech and Signal Processing}}, 2021,
  pp. 885--889.

\bibitem{Sato2021AmbisonicEquivariance}
R.~Sato, K.~Niwa, and K.~Kobayashi, ``{Ambisonic Signal Processing DNNs
  Guaranteeing Rotation, Scale and Time Translation Equivariance},''
  \emph{\protect\JournalTitle{IEEE/ACM Transactions on Audio Speech and
  Language Processing}}, vol.~29, pp. 1449--1462, 2021.

\bibitem{Phan2020OnLocalization}
H.~Phan, L.~Pham, P.~Koch, N.~Q.~K. Duong, I.~Mcloughlin, and A.~Mertins, ``{On
  Multitask Loss Function for Audio Event Detection and Localization},'' in
  \emph{\protect\JournalTitle{Proceedings of the 5th Workshop on Detection and
  Classification of Acoustic Scenes and Events}}, 2020.

\bibitem{Wang2021ADetection}
\BIBentryALTinterwordspacing
Q.~Wang, J.~Du, H.-X. Wu, J.~Pan, F.~Ma, and C.-H. Lee, ``{A Four-Stage Data
  Augmentation Approach to ResNet-Conformer Based Acoustic Modeling for Sound
  Event Localization and Detection},'' \emph{\protect\JournalTitle{arXiv}},
  2021.
\BIBentrySTDinterwordspacing

\bibitem{Shimada2021ACCDOA:Detection}
\BIBentryALTinterwordspacing
K.~Shimada, Y.~Koyama, N.~Takahashi, S.~Takahashi, and Y.~Mitsufuji, ``{ACCDOA:
  Activity-Coupled Cartesian Direction of Arrival Representation for Sound
  Event Localization And Detection},'' in
  \emph{\protect\JournalTitle{Proceedings of the IEEE International Conference
  on Acoustics, Speech and Signal Processing}}, 2021, pp. 915--919.

\bibitem{Lee2021SoundChallenge}
S.-H. Lee, J.-W. Hwang, S.-B. Seo, and H.-M. Park, ``{Sound Event Localization
  and Detection Using Cross-Modal Attention and Parameter Sharing for DCASE2021
  Challenge},'' Tech. Rep., 2021.

\bibitem{Xue2020SoundLearning}
W.~Xue, Y.~Tong, C.~Zhang, G.~Ding, X.~He, and B.~Zhou, ``{Sound event
  localization and detection based on multiple DOA beamforming and multi-task
  learning},'' in \emph{\protect\JournalTitle{Proceedings of the Annual
  Conference of the International Speech Communication Association}}, 2020, pp.
  5091--5095.

\bibitem{Park2020SoundFunctions}
S.~Park, S.~Suh, and Y.~Jeong, ``{Sound Event Localization and Detection with
  Various Loss Functions},'' Tech. Rep., 2020.

\bibitem{Emmanuel2021Multi-scaleDetection}
P.~Emmanuel, N.~Parrish, and M.~Horton, ``{Multi-scale Network for Sound Event
  Localization and Detection},'' Tech. Rep., 2021.

\bibitem{Kapka2019SoundModels}
S.~Kapka and M.~Lewandowski, ``{Sound Source Detection, Localization And
  Classification Using Consecutive Ensemble Of CRNN Models},'' Tech. Rep.,
  2019.

\bibitem{Wang2020TheChallenge}
Q.~Wang, H.~Wu, Z.~Jing, F.~Ma, Y.~Fang, Y.~Wang, T.~Chen, J.~Pan, J.~Du, and
  C.-H. Lee, ``{The USTC-iFlytek System for Sound Event Localization and
  Detection of DCASE2020 Challenge},'' Tech. Rep., 2020.

\bibitem{Shimada2021EnsembleDetection}
K.~Shimada, N.~Takahashi, Y.~Koyama, S.~Takahashi, E.~Tsunoo, M.~Takahashi, and
  Y.~Mitsufuji, ``{Ensemble of ACCDOA- and EINV2-based Systems with D3Nets and
  Impulse Response Simulation for Sound Event Localization and Detection},''
  Tech. Rep., 2021.

\bibitem{Politis2020ADetection}
A.~Politis, S.~Adavanne, and T.~Virtanen, ``{A Dataset of Reverberant Spatial
  Sound Scenes with Moving Sources for Sound Event Localization and
  Detection},'' in \emph{\protect\JournalTitle{Proceedings of the 5th Workshop
  on Detection and Classification of Acoustic Scenes and Events}}, 2020, pp.
  165--169.

\bibitem{Politis2021}
A.~Politis, S.~Adavanne, D.~Krause, A.~Deleforge, P.~Srivastava, and
  T.~Virtanen, ``{A Dataset of Dynamic Reverberant Sound Scenes with
  Directional Interferers for Sound Event Localization and Detection},'' in
  \emph{\protect\JournalTitle{Proceedings of the 6th Workshop on Detection and
  Classification of Acoustic Scenes and Events}}, 2021, pp. 125--129.

\bibitem{Adavanne2019ADetection}
S.~Adavanne, A.~Politis, and T.~Virtanen, ``{A Multi-room Reverberant Dataset
  for Sound Event Localization and Detection},'' in
  \emph{\protect\JournalTitle{Proceedings of the 4th Workshop on Detection and
  Classification of Acoustic Scenes and Events}}, 2019, pp. 10--14.

\bibitem{Nguyen2014RobustSources}
T.~N.~T. Nguyen, S.~K. Zhao, and D.~L. Jones, ``{Robust DOA estimation of
  multiple speech sources},'' in \emph{\protect\JournalTitle{Proceedings of the
  IEEE International Conference on Acoustics, Speech and Signal Processing}},
  2014, pp. 2287--2291.

\bibitem{Nguyen2020RobustNetwork}
T.~N.~T. Nguyen, W.~S. Gan, R.~Ranjan, and D.~L. Jones, ``{Robust Source
  Counting and DOA Estimation Using Spatial Pseudo-Spectrum and Convolutional
  Neural Network},'' \emph{\protect\JournalTitle{IEEE/ACM Transactions on Audio
  Speech and Language Processing}}, vol.~28, pp. 2626--2637, 2020.

\bibitem{Nguyen2021DCASEDetection}
T.~N.~T. Nguyen, K.~Watcharasupat, N.~K. Nguyen, D.~L. Jones, and W.~S. Gan,
  ``{DCASE 2021 Task 3: Spectrotemporally-aligned Features for Polyphonic Sound
  Event Localization and Detection},'' Tech. Rep., 2021.

\bibitem{Asano2007DetectionArray}
F.~Asano, K.~Yamamoto, J.~Ogata, M.~Yamada, and M.~Nakamura, ``{Detection and
  separation of speech events in meeting recordings using a microphone
  array},'' \emph{\protect\JournalTitle{EURASIP Journal on Audio, Speech, and
  Music Processing}}, vol. 2007, 2007.

\bibitem{Mohan2008LocalizationTest}
S.~Mohan, M.~E. Lockwood, M.~L. Kramer, and D.~L. Jones, ``{Localization of
  multiple acoustic sources with small arrays using a coherence test},''
  \emph{\protect\JournalTitle{Journal of the Acoustical Society of America}},
  vol. 123, no.~4, pp. 2136--2147, 2008.

\bibitem{Pavlidi2013Real-TimeArray}
D.~Pavlidi, A.~Griffin, M.~Puigt, and A.~Mouchtaris, ``{Real-Time Multiple
  Sound Source Localization and Counting Using a Circular Microphone Array},''
  \emph{\protect\JournalTitle{IEEE Transactions on Audio, Speech, and Language
  Processing}}, vol.~21, no.~10, pp. 2193--2206, 2013.

\bibitem{Zhao2015RobustReduction}
S.~Zhao, X.~Xiao, Z.~Zhang, T.~N.~T. Nguyen, X.~Zhong, B.~Ren, L.~Wang, D.~L.
  Jones, E.~S. Chng, and H.~Li, ``{Robust speech recognition using beamforming
  with adaptive microphone gains and multichannel noise reduction},'' in
  \emph{\protect\JournalTitle{Proceedings of the 2015 IEEE Workshop on
  Automatic Speech Recognition and Understanding}}, 2015, pp. 460--467.

\bibitem{Nguyen2020EnsembleTracking}
T.~N.~T. Nguyen, D.~L. Jones, and W.~S. Gan, ``{Ensemble of sequence matching
  networks for dynamic sound event localization, detection, and tracking},'' in
  \emph{\protect\JournalTitle{Proceedings of the 5th Workshop on Detection and
  Classification of Acoustic Scenes and Events}}, 2020, pp. 120--124.

\bibitem{Nguyen2019DCASEDetection}
T.~N.~T. Nguyen, D.~L. Jones, R.~Ranjan, S.~Jayabalan, and W.~S. Gan, ``{DCASE
  2019 Task 3: A two-step system for sound event localization and detection},''
  Tech. Rep., 2019.

\bibitem{Gerkmann2012UnbiasedDelay}
T.~Gerkmann and R.~C. Hendriks, ``{Unbiased MMSE-based noise power estimation
  with low complexity and low tracking delay},''
  \emph{\protect\JournalTitle{IEEE Transactions on Audio, Speech and Language
  Processing}}, vol.~20, no.~4, pp. 1383--1393, 2012.

\bibitem{Rafaely2017SpeakerStatistics}
B.~Rafaely and D.~Kolossa, ``{Speaker localization in reverberant rooms based
  on direct path dominance test statistics},'' in
  \emph{\protect\JournalTitle{Proceedings of the IEEE International Conference
  on Acoustics, Speech and Signal Processing}}, 2017, pp. 6120--6124.

\bibitem{Zhao2014UnderdeterminedSensor}
\BIBentryALTinterwordspacing
S.~Zhao, T.~Saluev, and D.~L. Jones, ``{Underdetermined direction of arrival
  estimation using acoustic vector sensor},''
  \emph{\protect\JournalTitle{Signal Processing}}, vol. 100, pp. 160--168,
  2014.
\BIBentrySTDinterwordspacing

\bibitem{Cao2019Two-StageCross-Correlation}
Y.~Cao, T.~Iqbal, Q.~Kong, M.~B. Galindo, W.~Wang, and M.~D. Plumbley,
  ``{Two-Stage Sound Event Localization and Detection using Intensity Vector
  and Generalized Cross-Correlation},'' Tech. Rep., 2019.

\bibitem{Delikaris-Manias2017DOAVectors}
S.~Delikaris-Manias, D.~Pavlidi, A.~Mouchtaris, and V.~Pulkki, ``{DOA
  estimation with histogram analysis of spatially constrained active intensity
  vectors},'' in \emph{\protect\JournalTitle{Proceedings of the 2017 IEEE
  International Conference on Acoustics, Speech and Signal Processing}}, 2017,
  pp. 526--530.

\bibitem{Kong2020PANNs:Recognition}
\BIBentryALTinterwordspacing
Q.~Kong, Y.~Cao, T.~Iqbal, Y.~Wang, W.~Wang, and M.~D. Plumbley, ``{PANNs:
  Large-Scale Pretrained Audio Neural Networks for Audio Pattern
  Recognition},'' \emph{\protect\JournalTitle{IEEE/ACM Transactions on Audio
  Speech and Language Processing}}, vol.~28, pp. 2880--2894, 2020.
\BIBentrySTDinterwordspacing

\bibitem{Adavanne2021DCASEInterference}
\BIBentryALTinterwordspacing
S.~Adavanne and A.~Politis, ``{DCASE 2021: Sound Event Localization and
  Detection with Directional Interference},'' 2021. [Online]. Available:
  \url{https://github.com/sharathadavanne/seld-dcase2021}
\BIBentrySTDinterwordspacing

\bibitem{Zhong2020RandomAugmentation}
Z.~Zhong, L.~Zheng, G.~Kang, S.~Li, and Y.~Yang, ``{Random erasing data
  augmentation},'' in \emph{\protect\JournalTitle{Proceedings of the 34th AAAI
  Conference on Artificial Intelligence}}, 2020, pp. 13\,001--13\,008.

\bibitem{Park2019SpecAugment:Recognition}
D.~S. Park, W.~Chan, Y.~Zhang, C.~C. Chiu, B.~Zoph, E.~D. Cubuk, and Q.~V. Le,
  ``{SpecAugment: A simple data augmentation method for automatic speech
  recognition},'' in \emph{\protect\JournalTitle{Proceedings of the Annual
  Conference of the International Speech Communication Association}}, 2019, pp.
  2613--2617.

\bibitem{Politis2020Overview2019}
A.~Politis, A.~Mesaros, S.~Adavanne, T.~Heittola, and T.~Virtanen, ``{Overview
  and Evaluation of Sound Event Localization and Detection in DCASE 2019},''
  \emph{\protect\JournalTitle{IEEE/ACM Transactions on Audio Speech and
  Language Processing}}, vol.~29, pp. 684--698, 2020.

\bibitem{Mesaros2019JointEvents}
A.~Mesaros, S.~Adavanne, A.~Politis, T.~Heittola, and T.~Virtanen, ``{Joint
  Measurement of Localization and Detection of Sound Events},'' in
  \emph{\protect\JournalTitle{Proceedings of the IEEE Workshop on Applications
  of Signal Processing to Audio and Acoustics}}, 2019.

\bibitem{Dmochowski2009OnArrays}
J.~Dmochowski, J.~Benesty, and S.~Aff{\`{e}}s, ``{On spatial aliasing in
  microphone arrays},'' \emph{\protect\JournalTitle{IEEE Transactions on Signal
  Processing}}, vol.~57, no.~4, pp. 1383--1395, 2009.

\end{thebibliography}

\clearpage
\begin{IEEEbiography}[{\includegraphics[width=1in,height=1.25in,clip,keepaspectratio]{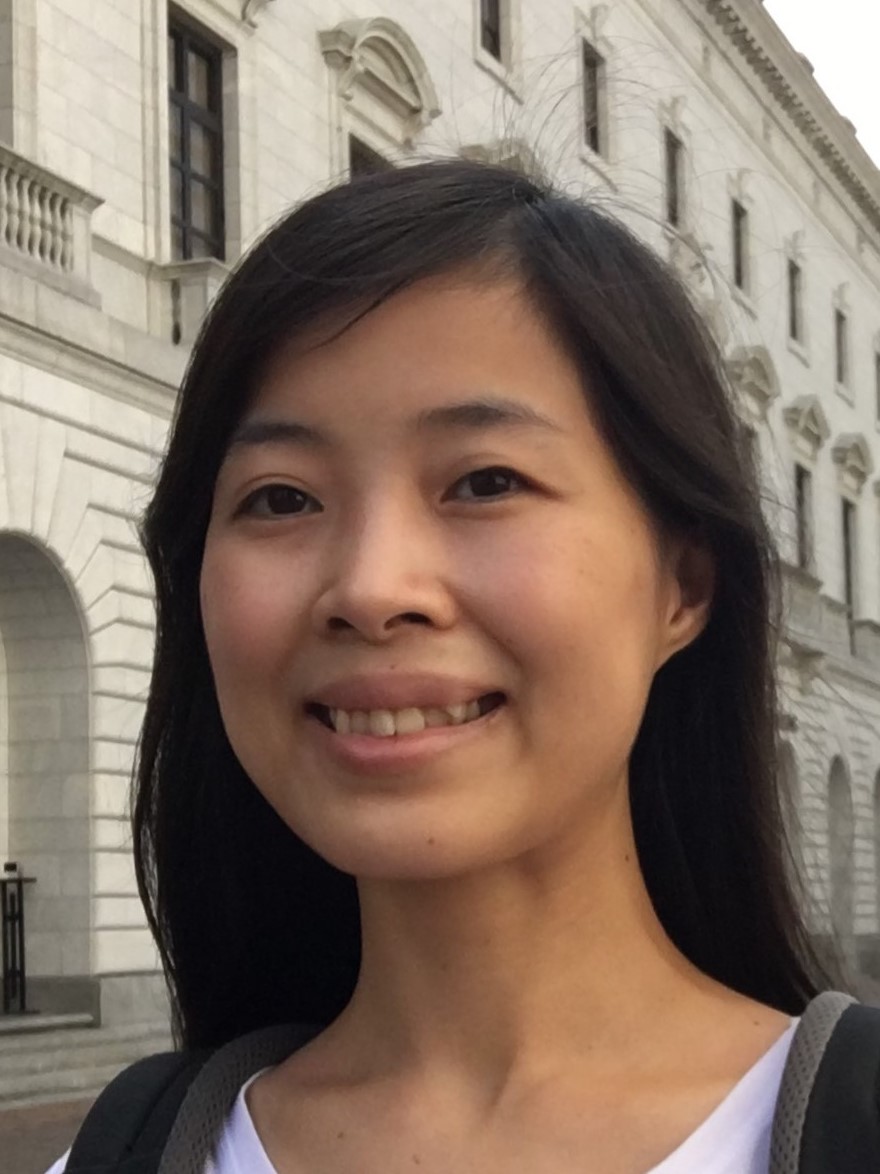}}]{Thi Ngoc Tho Nguyen} (S'19)
is a Ph.D. student at the Nanyang Technological University (NTU) in Singapore. Prior to joining NTU, she worked at University of Illinois Research Center in Singapore for five years as a research engineer. Her research interests are audio signal processing, deep learning, microphone array signal processing, and real-time processing. She has published several papers on the topics of direction-of-arrival estimation of multiple sound sources, and sound event localization and detection.
\end{IEEEbiography}

\begin{IEEEbiography}[{\includegraphics[width=1in,height=1.25in,clip,keepaspectratio]{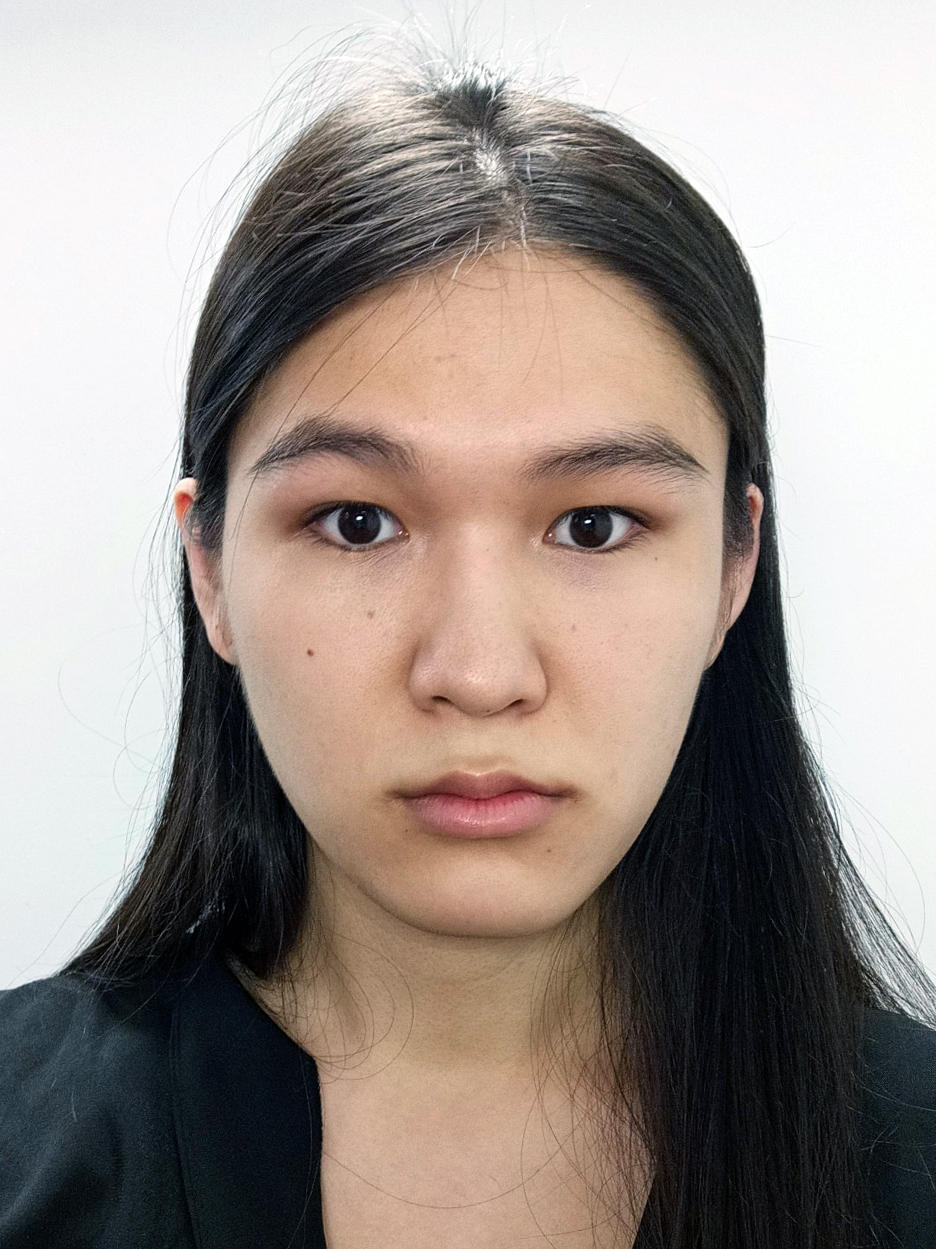}}]{Karn N. Watcharasupat} (S'19) was born in Bangkok, Thailand, in 1999. She received her B.Eng. (Hons) in Electrical and Electronic Engineering under the CN Yang Scholars Programme, from Nanyang Technological University (NTU), Singapore, in 2022. She is currently a research engineer at the School of Electrical and Electronic Engineering (EEE), NTU.

From 2018 to 2020, she was with the NTU EEE Media Technology Laboratory. In Spring 2020, she was a visiting research student at Music Informatics Group, Center for Music Technology (GTCMT), Georgia Institute of Technology, Atlanta, GA, USA, before returning again remotely since Spring 2021. Concurrently since 2021, she has been with the Digital Signal Processing Laboratory, the Smart Nation Translational Laboratory, and the Alibaba-NTU Singapore Joint Research Institute, NTU.

Her research interests are in signal processing, machine learning, and artificial intelligence for music and audio applications. Since 2021, she has published more than 10 papers in international conferences and journals on music information retrieval, soundscapes, spatial audio, speech enhancement, and blind source separation.
\end{IEEEbiography}

\begin{IEEEbiography}[{\includegraphics[width=1in,height=1.25in,clip,keepaspectratio]{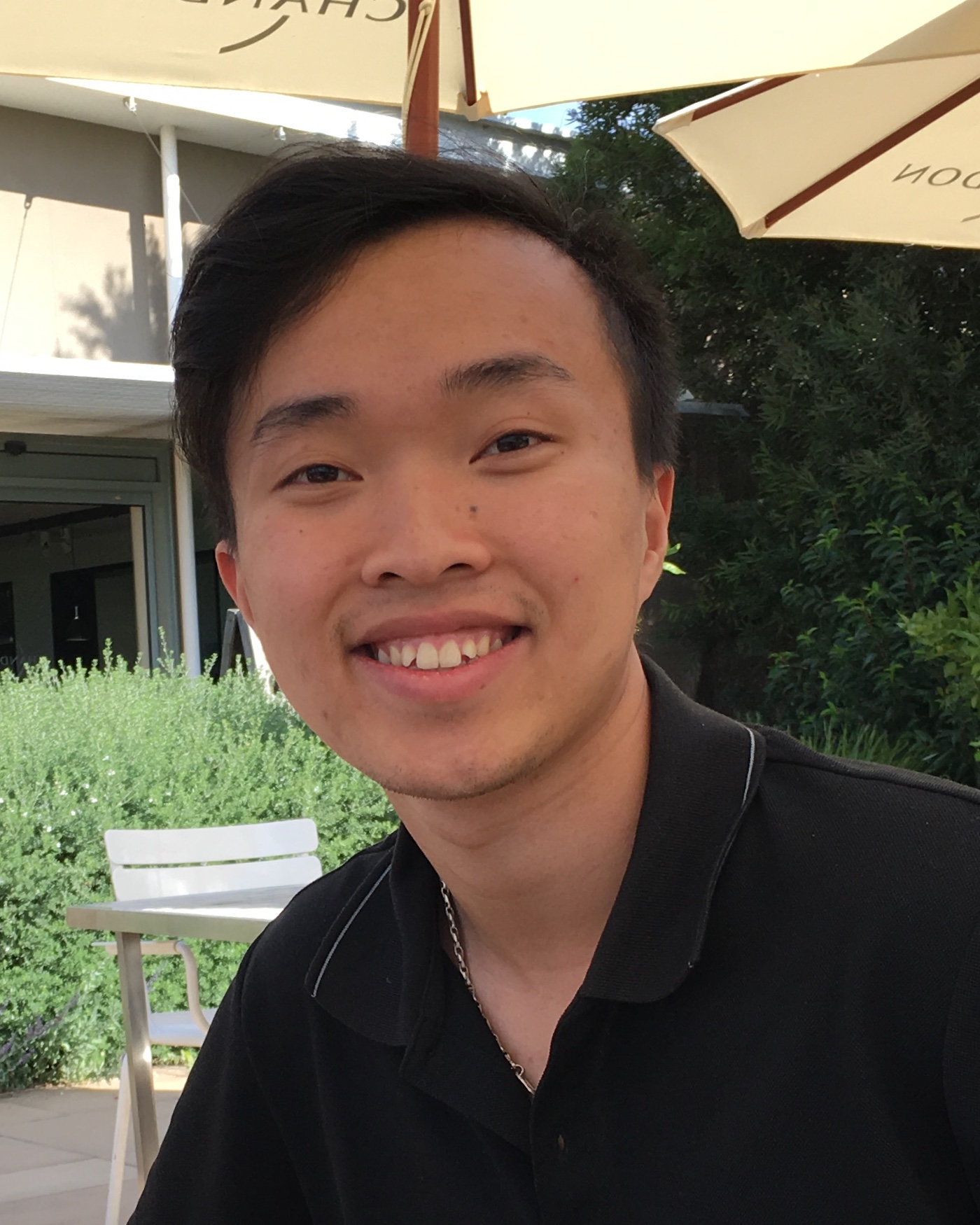}}]{Ngoc Khanh Nguyen} received his B.Eng. (Hons) in Electronics and Computer System from Swinburne University, Australia in 2017. He is currently a software engineer. He is keen on topics in computer sciences such as algorithms, database, machine learning and deep learning. He has also participated in several Kaggle competitions.
\end{IEEEbiography}

\begin{IEEEbiography}[{\includegraphics[width=1in,height=1.25in,clip,keepaspectratio]{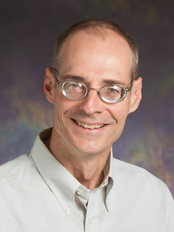}}]{Douglas L. Jones} (S’82–-M’83-–S’84–-M’87-–
SM’97–-F’02) received the BSEE, MSEE, and Ph.D. degrees from Rice University in 1983, 1986, and 1987, respectively. During the 1987-1988 academic year, he was at the University of Erlangen-Nuremberg in Germany on a Fulbright postdoctoral fellowship. Since 1988, he has been with the University of Illinois at Urbana-Champaign, where he is currently a Professor in Electrical and Computer Engineering, Neuroscience, the Coordinated Science Laboratory, and the Beckman Institute. He was on sabbatical leave at the University of Washington in Spring 1995 and at the University of California at Berkeley in Spring 2002. In the Spring semester of 1999 he served as the Texas Instruments Visiting Professor at Rice University. He is an author of two DSP laboratory textbooks, and was selected as the 2003 Connexions Author of the Year. He is a Fellow of the IEEE. He served on the Board of Governors of the IEEE Signal Processing Society from 2002-2004. His research interests are in digital signal processing, including nonstationary signal analysis, adaptive processing, multisensor data processing, OFDM, and various applications such as low-power implementations, biology and neuroengineering, and advanced hearing aids and other audio systems.  
\end{IEEEbiography}

\begin{IEEEbiography}[{\includegraphics[width=1in,height=1.25in,clip,keepaspectratio]{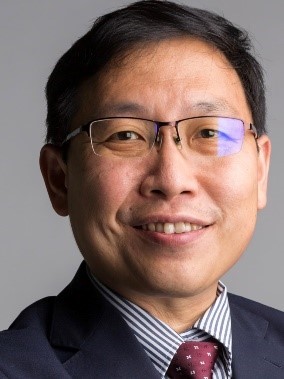}}]{Woon-Seng Gan} (S’90-–M’93–-SM’00)
received his BEng (1st Class Hons) and PhD degrees, both in Electrical and Electronic Engineering from the University of Strathclyde, UK in 1989 and 1993 respectively. He is currently a Professor of Audio Engineering and the Director of the Smart Nation Lab in the School of Electrical and Electronic Engineering in Nanyang Technological University. He also served as the Head of the Information Engineering Division in the School of Electrical and Electronic Engineering in Nanyang Technological University (2011-2014), and the Director of the Centre for Infocomm Technology (2016-2019). His research has been concerned with the connections between the physical world, signal processing and sound control, which resulted in the practical demonstration and licensing of spatial audio algorithms, directional sound beam, and active noise control for headphones and open windows.

He has published more than 400 international refereed journals and conferences, and has translated his research into 6 granted patents. He had co-authored three books on Subband Adaptive Filtering: Theory and Implementation (John Wiley, 2009); Embedded Signal Processing with the Micro Signal Architecture, (Wiley-IEEE, 2007); and Digital Signal Processors: Architectures, Implementations, and Applications (Prentice Hall, 2005). In 2017, he won the APSIPA Sadaoki Furui Prize Paper Award. He is a Fellow of the Audio Engineering Society (AES), a Fellow of the Institute of Engineering and Technology (IET), and a Senior Member of the IEEE. He served as an Associate Editor of the IEEE/ACM Transaction on Audio, Speech, and Language Processing (TASLP; 2012-15) and was presented with an Outstanding TASLP Editorial Board Service Award in 2016. He also served as the Associate Editor for the IEEE Signal Processing Letters (2015-19). He is currently serving as a Senior Area Editor of the IEEE Signal Processing Letters (2019-); Associate Technical Editor of the Journal of Audio Engineering Society (JAES; 2013-); Editorial member of the Asia Pacific Signal and Information Processing Association (APSIPA; 2011-) Transaction on Signal and Information Processing; Associate Editor of the EURASIP Journal on Audio, Speech and Music Processing (2007-).
\end{IEEEbiography}

\end{document}